\documentclass[10pt,journal,compsoc]{IEEEtran}
\ifCLASSOPTIONcompsoc
\usepackage[nocompress]{cite}
\else
\usepackage{cite}
\fi

\ifCLASSINFOpdf
\usepackage[pdftex]{graphicx}
\else
\fi

\usepackage{algorithmic}
\usepackage{algorithm}

\usepackage{amsmath}
\usepackage{graphicx}
\usepackage{epstopdf} 

\usepackage{lipsum}
\usepackage{amssymb,amsmath,amsthm}
\usepackage[normalem]{ulem}

\newtheorem{problem}{Problem}

\usepackage{mathtools} 

\usepackage{multirow}

\usepackage[table,xcdraw]{xcolor}

\usepackage{booktabs}

\usepackage[flushleft]{threeparttable}
\usepackage{pdfpages}
\usepackage{makecell}

\usepackage{fancyhdr}

\fancypagestyle{firstpage}{%
	\lhead{Under review}
	\rhead{}
}
\title{Problem Set 1}

\begin{document}
	%
	\title{TEAGS: Time-aware Text Embedding Approach to Generate Subgraphs}
	%
	%
	%
	%

	\author{Saeid~Hosseini,
		Saeed~Najafipour,        
		Ngai-Man~Cheung,
		Hongzhi~Yin,
		Mohammad~Reza~Kangavari, and
		Xiaofang~Zhou		        
		\IEEEcompsocitemizethanks{\IEEEcompsocthanksitem S. Hosseini and N. Cheung are with the ST Electronics - SUTD Cyber Security Lab., Singapore University of Technology and Design, Singapore.\protect\\Email: ngaiman\_cheung@sutd.edu.sg, saeid.hosseini@uq.net.au
			\IEEEcompsocthanksitem Xiaofang Zhou and Hongzhi Yin are with School of Information Technology and Electrical Engineering, University of Queensland, Brisbane, Australia.\protect\\Email: \{zxf,h.yin1\}@itee.uq.edu.au
			\IEEEcompsocthanksitem S. Najafi Pour, M. R. Kangavari, and S.Hosseini(Affiliated) are with computational cognitive model research lab., School of computer engineering, Iran University of Science and Technology, Iran. \protect\\Email: {saeed\_najafi}@comp.iust.ac.ir, {kanagvari}@iust.ac.ir
		}
		\thanks{}}
	
	%
	%
	
	\markboth{}%
	{Shell \MakeLowercase{\textit{et al.}}: Bare Demo of IEEEtran.cls for Computer Society Journals}
	%
	
	
	
	\IEEEtitleabstractindextext{%
		\begin{abstract}
			Given a graph over which the contagions (e.g. virus, gossip) propagate, leveraging subgraphs with highly correlated nodes
			is beneficial to many applications. Yet, challenges abound. First, the propagation pattern between a pair of nodes may change in
			various temporal dimensions. 
			Second, not always the same contagion is propagated. Hence, state-of-the-art text mining
			approaches ranging from similarity measures to topic-modeling cannot use the textual contents to compute the weights between the nodes.
			Third, the word-word co-occurrence patterns may differ in various temporal dimensions, which increases the difficulty to employ current
			word embedding approaches. We argue that inseparable multi-aspect temporal collaborations are inevitably needed to better calculate the correlation metrics
			in dynamical processes. In this work, we showcase a sophisticated framework that on the one hand, integrates a neural network based time-aware word
			embedding component that can collectively construct the word vectors through an assembly of infinite latent temporal facets, and on the other hand, uses an elaborate generative model to compute the edge weights through heterogeneous temporal attributes. After computing the intra-nodes weights, we utilize our Max-Heap Graph cutting algorithm to exploit subgraphs. We then validate our model
			through comprehensive experiments on real-world propagation data. The results show that the knowledge gained from the versatile
			temporal dynamics is not only indispensable for word embedding approaches but also plays a significant role in the understanding of the
			propagation behaviors. Finally, we demonstrate that compared with other rivals, our model can dominantly exploit the subgraphs with highly coordinated nodes.
		\end{abstract}
		\vspace{-3mm}
		}		
		\maketitle
\thispagestyle{firstpage}

		\IEEEpeerreviewmaketitle
		\IEEEraisesectionheading{\section{Introduction}\label{sec:introduction}}
		\IEEEPARstart{P}{ropagation} process over a graph of vertices is

		multi-disciplinary and involves with epidemiology \cite{Zhang2015}, viral vigilante \cite{Sepehr2018} \cite{Zhang2018}, social media \cite{Peng2017}, and cybersecurity \cite{hosseini2018mining}. Given a graph $G$ as a propagation network over which the contagions (e.g. gossip, virus, and alarm) spread, a subgraph $g$ of $G$ is tightly connected if all the nodes in $g$ are highly correlated with each other. Identifying tightly connected subgraphs finds important applications in numerous domains. For example, highly correlated subgraphs are used (i) in transport systems to reorder the repairs and reduce the number of future defects \cite{hosseini2018mining}\cite{Ni2017}, (ii) in epidemiology to vaccinate influential nodes and prevent the spread of the disease \cite{Medlock2009}, and (iii) in social media to manipulate (add or delete) the network edges and control outspreading of the gossips \cite{chen2016node}\cite{Khalil2014}. Moreover, tightly connected subgraphs can be beneficial for pre-emptive schemes such as preventive medicine and preventative maintenance.\\
		Numerous techniques \cite{Peng2017}\cite{GomezRodriguez2010}\cite{Goyal2010}\cite{Prakash2010} have been proposed to compute the edge weights among each pair of the nodes in propagation networks that, roughly speaking, is the main step in exploiting of the subgraphs. This reduces the NP-hard problem of subgraph mining to \textit{edge weight calculation} that is tailored by a \textit{Graph-cut algorithm}. Although, the computing of the edge weights is conceptually simple, significant obstacles incurred by the specifications of the propagation process. Therefore, new approaches must be introduced to handle them. In the following, we present the \textit{challenges} to illustrate such a need.\\
		\textit{\textbf{Challenge 1 (Varied Temporal Pattern)}} \\
		\indent The propagation pattern between a pair of nodes ($n_i,n_j$) may differ in \textit{various latent temporal facets} (e.g. hour, day) that are revealed as parts of the same whole. To exemplify the hour dimension, while the node $n_i$ may spread a contagion to $n_j$ during the day, the propagation may decay to some extent or cease throughout the night hours. The polymorphous \textit{depth} feature even makes hour factor dependent on parent facets (i.e. day and week).
		However, the traditional remarkable models in inferring of the propagation processes \cite{Peng2017}\cite{Prakash2010}\cite{GomezRodriguez2010}\cite{Goyal2010} only consider the \textit{sequential} interlude attribute. The rule seems to be that the farther apart in time the two nodes spread a contagion in between, the less likely they are to collaborate tightly.\\
		\textit{\textbf{Challenge 2 (Diverse Propagation Direction)}}\\
		\indent Not only the type of contagion and the amount of the functional load can differ in various temporal facets, but owing to the temporal dimension specifications, the \textit{propagation direction} may also alter. e.g., While the node $n_i$ influences node $n_j$ from Monday to Friday, the influence can either be eliminated or divert inversely during the weekends.\\
		\textit{\textbf{Challenge 3 (Mismatched Node Contents)}} \\
		\indent Unlike prior work \cite{Peng2017}\cite{GomezRodriguez2010}\cite{Goyal2010}\cite{Prakash2010}, not always the same contagion is propagated \cite{Zhang2017} to the subsequent influenced node. For instance, an \textit{oil leakage} initially causes an \textit{extra vibration}, continues with \textit{overheating}, and finishes with \textit{engine failure}. As a result, \textit{node-specific textual contents} in a propagation sequence may end up \textit{mismatched}. Consequently, the textual contents associated with the pair of nodes will not gain sufficient statistical signals to utilize current text mining approaches (e.g., topic modeling \cite{nguyen2015improving} and other heuristics \cite{Hosseini2014}). As a result, the correlation edge weight between the node pair will not be computed appropriately.\\
		\textit{\textbf{Challenge 4 (Temporally Skewed Contents)}} \\
		\indent Given the corpus $\mathcal{C}$ comprising the textual contents of the nodes, the co-occurrence matrix $X$ reports how each pair of words co-occurs in $\mathcal{C}$. The co-occurrence matrix is the essential element for vector representation. However, current state-of-the-art word vector representation models \cite{dumais2004latent}\cite{Mikolov2013a} neglect the fact that the \textit{word proximities} can be affected by the various latent temporal facets.
		\par \textbf{\textit{Observation:}} In order to study how the word co-occurrence patterns change in \textit{various temporal dimensions}, we set up an observation on a dataset \cite{Hosseini2014} of one million tweets in Australia. Tables \ref{Word-probability-Hour-dimension} and \ref{Word-probability-season-dimension} show the distribution probability of the word co-occurrences in the hour and season latent factors. As Table \ref{Word-probability-Hour-dimension} demonstrates, almost one-third of the times the word \textit{go} co-occurs with \textit{work} during 6 to 11 am. Compared to those who start the work early in the morning, the night shift people tend more to \textit{drive} to \textit{work} in the evening. Moreover, from the co-occurrence of \textit{Have+Tea}, we understand that more people have the tea in the morning. Finally, people \textit{have+fun} more during 6-11 am that is affected by the weekend tweets.	
		\vspace{-3mm}
		\begin{table}[H]	
			\caption{Co-occurrence probability - Hour dimension}
			\vspace{-2mm}
			\centering
			\begin{tabular}{|c|c|c|c|c|c|}
				\hline
				Word 1 & Word 2 & 0 - 5                         & 6 - 11                        & 12 - 17 & 18 - 23                       \\ \hline
				Go     & Work   & 0.295                         & \textbf{0.324} & 0.119   & \textbf{0.260} \\ \hline
				Drive  & Work   & 0.144                         & \textbf{0.289} & 0.108   & \textbf{0.457} \\ \hline
				Have   & Tea    & \textbf{0.303} & \textbf{0.383} & 0.147   & 0.165                         \\ \hline
				Have   & Fun    & \textbf{0.306} & \textbf{0.388} & 0.143   & 0.161                         \\ \hline
			\end{tabular}
			\label{Word-probability-Hour-dimension}
			\vspace{-2mm}
		\end{table}
		\vspace{-3mm}
		 \par Also as Table \ref{Word-probability-season-dimension} reports, the word \textit{cold} mainly comes with \textit{drink} during the summer and the people dispute about the \textit{hot+day} and \textit{hot+night} mostly during the spring and summer. Also, the distribution of \textit{go+surf} exhibits that summer and fall are respectively the most popular seasons for surfing. Thus, the word pair co-occurrence patterns are temporally skewed and conceivably differ in various dimensions.
		 \vspace{-5mm}
		\begin{table}[H]
			\caption{Co-occurrence probability - Season dimension}
			\vspace{-2mm}
			\centering
			\begin{tabular}{|c|c|c|c|c|c|}
				\hline
				Word 1 & Word 2 & Spring                        & Summer                        & Fall                          & Winter \\ \hline
				Cold   & Drink  & 0.228                         & \textbf{0.485} & 0.142                         & 0.142  \\ \hline
				Hot    & Day    & \textbf{0.265} & \textbf{0.371} & 0.220                         & 0.142  \\ \hline
				Hot    & Night  & \textbf{0.269} & \textbf{0.387} & 0.203                         & 0.139  \\ \hline
				Go     & Surf   & 0.215                         & \textbf{0.268} & \textbf{0.364} & 0.151  \\ \hline
			\end{tabular}
			\vspace{-3mm}
			\label{Word-probability-season-dimension}
		\end{table}
		\par
		\textbf{\textit{Contributions.}} In this work, revealing the \textit{utopian spirit of the time}, we state that the \textit{multi-aspect temporal knowledge} \cite{hosseini2017leveraging} of the propagation network is indispensable for subgraph mining. This in turn benefits
		many real-world applications that deal with the spread of the contagions in various domains. According to the above discussion, three types
		of requirements are proposed to cope with the challenges in exploiting of the tightly connected subgraphs: 1) providing a better insight into the word co-occurrence patterns in multiple dimensions; 2) comprising various attributes to better understand discrete temporal dynamics \cite{Abachi2018}; 3) appraising propagations in both directions; Such needs are met in Sections \ref{time-aware-word-embedding}, \ref{Multifacet-Generative-Model}, and \ref{Scoring-Propagation-Coherence}.\\	To the best of our knowledge, we propose the first study on multi-aspect time-aware word embedding which aims to leverage the subgraphs from bidirectional propagation networks. Our contributions in this work are threefold:
		\vspace{-2mm}
		\begin{itemize}
			\item We develop a neural network based time-aware word embedding approach that can collectively construct the word vectors based on an infinite number of temporal dimensions.
		    \item We devise a generative model that can compute the bilateral correlation weight between each pair of nodes through multi-aspect latent temporal facets.
			\item We design a Heap-based graph-cutting algorithm to maximize the correlation inside each of exploited subgraphs and simultaneously minimize the number of utilized edges in subgraph mining procedure.
			\item We propose a temporal-textual framework that can achieve better effectiveness in leveraging of the tightly connected subgraphs from bidirectional propagation networks.
		\end{itemize}
		\vspace{-7mm}
		\subsection{Paper Organization}
		\vspace{-1mm}
		\indent We organize the rest of this paper as follows: in Sec. \ref{Related_Work}, we briefly study the literature; in Sec. \ref{Problem Statement}, we clarify the problem for subgraph mining in bilateral propagation networks and provide a framework overview; our approaches and experiments are described in Sec. \ref{Methodology} and Sec. \ref{Experiments} respectively. Finally, we conclude this paper and discuss future work in Sec. \ref{Conclusion}.
		\vspace{-5mm}
		\section{Related Work}
		\label{Related_Work}
		\vspace{-1mm}
		As we propose a time-aware word-embedding approach to infer propagation processes, the literature includes \textit{Word embedding}, \textit{Propagation inference}, and \textit{Immunization policies}.
		\vspace{-4mm}
		\subsection{Word Embedding}
		\vspace{-1mm}
		\textit{Word embedding} exploits a real-valued vector for each given word in the corpus. Such vectors can better discover the correlation between the words and tackle relevant challenges in various domains such as Natural Language Processing (NLP) \cite{ling2015two}, Information Retrieval (IR) \cite{Ebraheem2018}, and Recommendation Systems (RS) \cite{park2018conceptvector} \cite{Hosseini2017a}. Traditional embedding methods include \textit{Bag-of-Words}(BOW) \cite{Manning2008} and \textit{Latent Semantic Indexing} (LSI) \cite{Deerwester1990}. BOW represents each document with its comprising words and while neglecting the order of vocabularies, relies on word frequencies. Conversely, LSI applies \textit{Singular Value Decomposition}(SVD) to identify the patterns between terms and concepts. Moreover, the BOW model \cite{ling2015not} can be extended using Expectation-Maximization(EM) \cite{talley2011database} to eliminate the word ambiguity, which particularly benefits machine translation. Furthermore, the state-of-the-art Neural word embedding models (e.g. \textit{Word2vec} \cite{Mikolov2013a} and \textit{GloVe} \cite{Pennington2014}) extract meaningful syntactical and semantical regularities from word pairs.		
Similar to BOW, the word embedding approaches can further be equipped with other algorithms \cite{nguyen2015improving}, \cite{zhu2018exploiting}, and \cite{Maskouni2018}. For instance, Nguyen et al. \cite{nguyen2015improving} fuses \textit{topic models} with embedding approaches to collectively generate a word either from Dirichlet multinominal component or from the continuous embedding module. Likewise, aiming to perform entity disambiguation, Zhu et al. \cite{zhu2018exploiting} enrich the word embedding modules with Knowledge Graphs (KG) to make the best utilization of both knowledge and corpus-based semantic similarities. 
Nevertheless, vector representation models are not only proposed for words \cite{Mikolov2013a}, but also involve other data structures, like Concept-Vector \cite{park2018conceptvector}. 
In general, the co-occurrence matrix is the essential element in \textit{count-based representation} methods. However, the word-word co-occurrence patterns are temporally skewed. 
Yet, many word embedding approaches \cite{Mikolov2013a} \cite{dumais2004latent} \cite{Pennington2014} neglect the temporal dynamics. On the contrary, the time-oriented models \cite{bamler2017dynamic} \cite{dubossarsky2017outta} utilize terminology in a single dimension to address various use-cases, including temporal word analogies, word-to-word relatedness \cite{rosin2017learning}, and semantical relations \cite{rosin2017learning}. Nonetheless, the temporal word pair co-occurrences may alter in \textit{diverse latent temporal facets} (e.g. hours, days, weeks, and month). Hence, in this paper, we devise a novel multifaceted time-aware word embedding approach which can conjointly infer word vectors through multiple temporal dimensions.		
		\vspace{-3mm}
		\begin{table}[h]
			\centering
			\caption{Literature}
			\vspace{-3mm}
			\def\arraystretch{1.5}
			\begin{tabular}{ c l l }
				
				\Xhline{2\arrayrulewidth}
				\vspace{1mm}
				Category                                                                         & Approach            & References \\ \hline
				& Traditional Models  &  \cite{Manning2008}\cite{Deerwester1990}\cite{ling2015not}\cite{talley2011database}         \\ 
				\multirow{2}{*}{\begin{tabular}[c]{@{}c@{}}Word\\ Embedding\end{tabular}}        & Neural Network      & \cite{Mikolov2013a}\cite{Pennington2014}\cite{Ebraheem2018}           \\ 
				& Hybrid              & \cite{nguyen2015improving}\cite{park2018conceptvector}\cite{zhu2018exploiting}           \\ 
				& Time-oriented       & \cite{bamler2017dynamic}\cite{dubossarsky2017outta}\cite{rosin2017learning}           \\ \hline
				
				& Cascade            & \cite{Prakash2012a}\cite{Hethcote2000}           \\ 
				\begin{tabular}[c]{@{}c@{}}Propagation\\ Inference\end{tabular}                & Threshold          & \cite{Khalil2014}\cite{Kempe2003}\cite{Ganesh2005}           \\  
				& Time-oriented      & \cite{GomezRodriguez2010}\cite{Goyal2010}\cite{hosseini2017leveraging}           \\ \hline
				& Single Node         & \cite{chen2016node}\cite{Khalil2014}\cite{Cohen2003}\cite{Saha2015}           \\  
				\multirow{2}{*}{\begin{tabular}[c]{@{}c@{}}Immunization\\ Policy\end{tabular}}   & Group Based         & \cite{hosseini2018mining}\cite{Hosseini2018}\cite{Zhang2015}\cite{Medlock2009}\cite{Peng2017}
				\\ 
				& Preventative        & \cite{Hosseini2018}           \\ 
				& Time-oriented       & \cite{GomezRodriguez2010}\cite{Goyal2010}           \\ \Xhline{2\arrayrulewidth}				
			\end{tabular}
			\vspace{-3mm}
		\end{table}
		\vspace{-3mm}
		\subsection{Propagation Inference}
		\vspace{-1mm}
		A \textit{Diffusion} or \textit{propagation} process interprets how each contagion (e.g. alarm or virus) may spread in a graph. The conventional models suppose that the nodes spread the contagions equally \cite{Anderson1992}. In general, the propagation inference algorithms are of two types of \textit{cascade} \cite{Prakash2012a}
		 and \textit{threshold-based} \cite{Yoo2017} \cite{Ni2017} \cite{Kempe2003} \cite{Khalil2014}. The threshold based approaches like \textit{Linear Threshold} (LT) determine under what verge the contagion may outspread. Specifically, \cite{Ganesh2005} claim the epidemic threshold as the first eigenvalue of the \textit{adjacency matrix} that is associated with the diffusion graph. On the contrary, the cascade models examine the condition of each vertex during propagation phase: \textit{Susceptible}, \textit{Infected}, \textit{Removed}, or \textit{Vigilant}. Subsequently, depending on the status of each vertex, one can track how the contagions are transmitted. From immunity perspective the cascade models are divided into three categories \cite{Hethcote2000}: without immunity (\textit{$SIS$}), provisional immunity (\textit{$SIRS$}), and perpetual immunity (\textit{$SIR$}). 
		 		 Though, newer approaches amend either the number of contagions (\textit{$SI_1I_2R$}) or the subgraph structure, e.g. \textit{clique}, \textit{chain}, and \textit{star}.
		 Furthermore, the contagions may spread in either or both directions \cite{GomezRodriguez2010}.\\
		 For the time aspect, the retrospective \textit{Sequential} attribute \cite{GomezRodriguez2010} \cite{Goyal2010} elucidates that the farther apart in-time a pair of vertices spread a contagion, the less correlation weight will be assigned to the edge between them. However, we argue that the dynamical processes may vary in different temporal dimensions. Hence, we need to collectively infer propagations in various dimensions. Owing to the fact that the vector representation of the words in textual contents can be distorted by the \textit{temporal permutations} \cite{hosseini2017leveraging}, we recommend a diligent neural network based \textit{multi-aspect time-aware embedding} module to mutually construct the word vectors through an infinite number of temporal dimensions.		
		\vspace{-6mm}
		\subsection{Immunization Policies}
		\vspace{-1mm}
		Immunization aims to stop or slow down the spread of contagions. \textit{Vaccination} \cite{Prakash2013} and \textit{social distancing} \cite{Shim2013} are the two common immunization tactics. Specifically, immunization can be applied on \textit{individual} or \textit{group} (i.e. subgraphs) of vertices. Single node models include \textit{random vaccination} \cite{Cohen2003}, \textit{network manipulation} \cite{Khalil2014,chen2016node}, \textit{data-driven} \cite{Zhang2017}, and recently \textit{non-spectral models} \cite{Saha2015} \cite{Khalil2014}. 		
		Furthermore, even though leveraging of the subgraphs typically result in sub-optimal solutions and the exploited subgraphs may also reshape \cite{Hosseini2018}, the group-based inoculation policies \cite{Zhang2016} can better control the outspread.
		Nevertheless, our proposed research benefits immunization tactics in three ways: First, where prior works \cite{GomezRodriguez2010} \cite{Goyal2010} only consider the sequential temporal property, we devise a temporal-textual approach which infers the edge weights via miscellaneous time-oriented features. Second, our group-based approach can better restrain dynamical procedures \cite{Zhang2016}. Third, unlike limited directed networks \cite{Peng2017}, we track intra-nodes propagations bilaterally.
		\vspace{-4mm}
\section{Problem Statement}
		\label{Problem Statement}
\vspace{-1mm}
		In this section, we define the concepts and problems of the paper. We then explain our framework overview.
		\vspace{-3mm}
		\subsection{Preliminary Concepts}
				\vspace{-1mm}
		The preliminary definitions are introduced as follows:
		\vspace{-2mm}
		\subsubsection{Definition 1 (node)} 
		A Node with identifier $n_i$ $\in$ $\mathbb{N}$ is a vertex over
		which the contagions propagate.
		\vspace{-2mm}
		\subsubsection{Definition 2 (alarm)}
An alarm $a_i \in$ $\mathbb{A}$ is recorded when a node gets infected by a contagion. Each alarm has an identity ($a_i$), pertinent node $n_m$, \textbf{c}ategory $c_l$, infection \textbf{t}ime $a_i$.t, and \textbf{s}tring contents $a_i.s$. $\mathbb{A}$ represents the alarm sets associated with each of the nodes ($\mathbb{A} = \lbrace A_1, A_2,...., A_n\rbrace$). Hence, $A_i$ denotes the set of times when the node $n_i$ has raised any of the alarms. 
As far as the type of the contagions is the same, the propagation can be admitted. In a \textit{medical context}, influenza can spread in three forms of $A$, $B$, and $C$, but all are counted as the same virus.
		\vspace{-6mm}
		\subsubsection{Definition 3 (propagation network)}
		\vspace{-1mm}		
		The undirected graph $G = (\mathbb{N},\mathbb{L})$ with the set of nodes $\mathbb{N}$ and links $\mathbb{L}$ comprises numerous interleaved \textit{directed acyclic graphs} that reflect how contagions spread bidirectionally.
		\vspace{-2mm}
		\subsubsection{Definition 4 (latent temporal facets)}
		\vspace{-1mm}
		The propagation behavior between a given pair of nodes ($n_i, n_j$) can differ based on the set of \textit{latent temporal facets}: $\mathbb{T} = \lbrace
		z^1,z^2,...,z^t\rbrace$. While our proposed solution (Section \ref{Multifacet-Generative-Model}) can accommodate an infinite number of temporal dimensions, owing to density we limit $\mathbb{T}$ to four dimensions of hour, day, week and month $\mathbb{T} = \lbrace z^h,z^d,z^w,z^m \rbrace$
		\vspace{-2mm}
		\subsubsection{Definition 5 (unifacet temporal slab)}
		\vspace{-1mm}
		Each latent temporal dimension $z^x$ can comprise $\eta$ splits $z^x = \lbrace s_{1}^{x},s_{2}^{x},...,s_{\eta}^{x} \rbrace$, e.g. 12 splits for month dimension. Accordingly, the \textit{Unifacet Temporal Slabs} are built via merging of the similar splits.
		\vspace{-3mm}
		\subsubsection{Definition 6 (multifacet temporal slab)}
		\vspace{-1mm}
		Given a set of
		unifacet temporal slabs belonging to $t$ hierarchical temporal dimensions, a \textit{Multifacet Temporal Slab} is designated by a
		combination of $b$ unifacet temporal slabs.
		\vspace{-3mm}
		\subsubsection{Definition 7 (vector representation)}
		\vspace{-1mm}
		Given the corpus of textual contents $\mathcal{C}$, the word embedding can convey a real-valued vector $\vec{v_i}$ to every vocabulary $v_i \in \mathbb{V}$. Here $\vec{v_i}$ is the vector which represents the word $v_i$.
		\vspace{-4mm}
		\begin{figure}[h]
			\centering
			\includegraphics[width=0.30\textwidth]{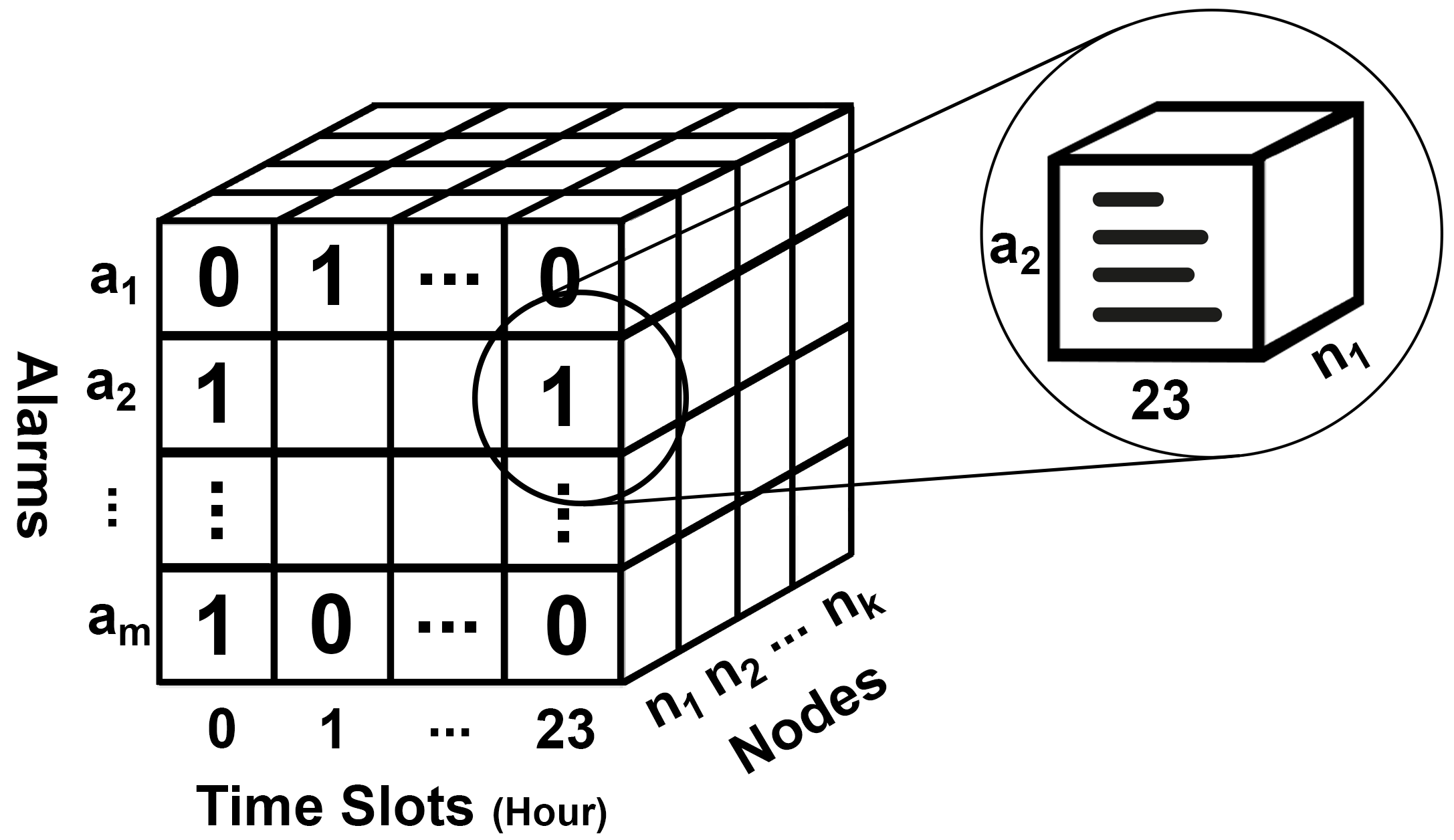}
			\vspace{-2mm}
			\caption{$\mathbb{NTA}$ Cube}
			\label{fig:cube}
		\end{figure}
		\vspace{-9mm}
\subsection{Problem Definition}
		The notations $\mathbb{N}$, $\mathbb{T}$, and $\mathbb{A}$ are respectively the set of \textbf{N}odes, \textbf{T}emporal facets, and \textbf{A}larms. Given a temporal dimension, the alarm time-stamps including the textual contents can be recorded in the dimension-specific $\mathbb{NTA}$ cube. Fig. \ref{fig:cube} illustrates the hour dimension cube, where each cell includes the possible textual contents of the alarms and the binary value to signify whether the sample node $n_i$ has raised the alarm $a_j$ at the particular temporal split or not. Since each node can get infected by a limited number of contagions, the $\mathbb{NTA}$ cube can be extremely sparse. 
		Consequently, we can put a \textit{proposition} forward: \textit{if two splits in a dimension are highly similar, we can predict unforeseen propagations in one of them based on the data recorded for the other.} Hence, we can demote the sparsity through merging of the similar splits.\\
 The problems addressed in this paper are as follows:
		\vspace{-2mm}
		\begin{problem}
			\label{extracting-multifacet-temporal-slabs}
			(extracting multifacet temporal slabs) Given the set of nodes $\mathbb{N}$, alarms $\mathbb{A}$, and the temporal dimensions $\mathbb{T}$, our goal is to extract all multifacet temporal slabs through merging highly similar splits in each of $t$ temporal dimensions.
		\end{problem}	
		\vspace{-4mm}
		\begin{problem} (mining correlation weights)
			Given a set of temporal dimensions $\mathbb{T}$ and the set of multifacet temporal slabs $\tau^b$, our aim is to compute the edge weight between each pair of nodes ($n_i$,$n_j$) using both temporal and textual data in propagation network.
			\label{Probtemporalslabs}	
		\end{problem}
		\vspace{-4mm}
		\begin{problem}
			(exploiting highly correlated subgraphs) Given a
			bilateral network $G$ over which the alarms ($\mathbb{A}$) spread, we aim to mine a set of subgraphs ($\mathbb{S}$), where the nodes in each of exploited subgraphs are highly correlated.
			\label{mininghighlycorrelatedsubgraphs}
		\end{problem}
		\vspace{-2mm}
		Intuitively, the problem of subgraph mining (Prob. \ref{mininghighlycorrelatedsubgraphs}) can be divided into two steps: Firstly, to calculate the edge weights between nodes (Prob. \ref{Probtemporalslabs}). Secondly, to employ a graph-cut algorithm to optimize the number of exploited subgraphs and maximize the intra-subgraph correlations. 
		\vspace{-5mm}
		\subsection{Framework Overview}
		\label{Framework-Overview}
\begin{figure}[H]
					\centering
					\includegraphics[width=0.49\textwidth]{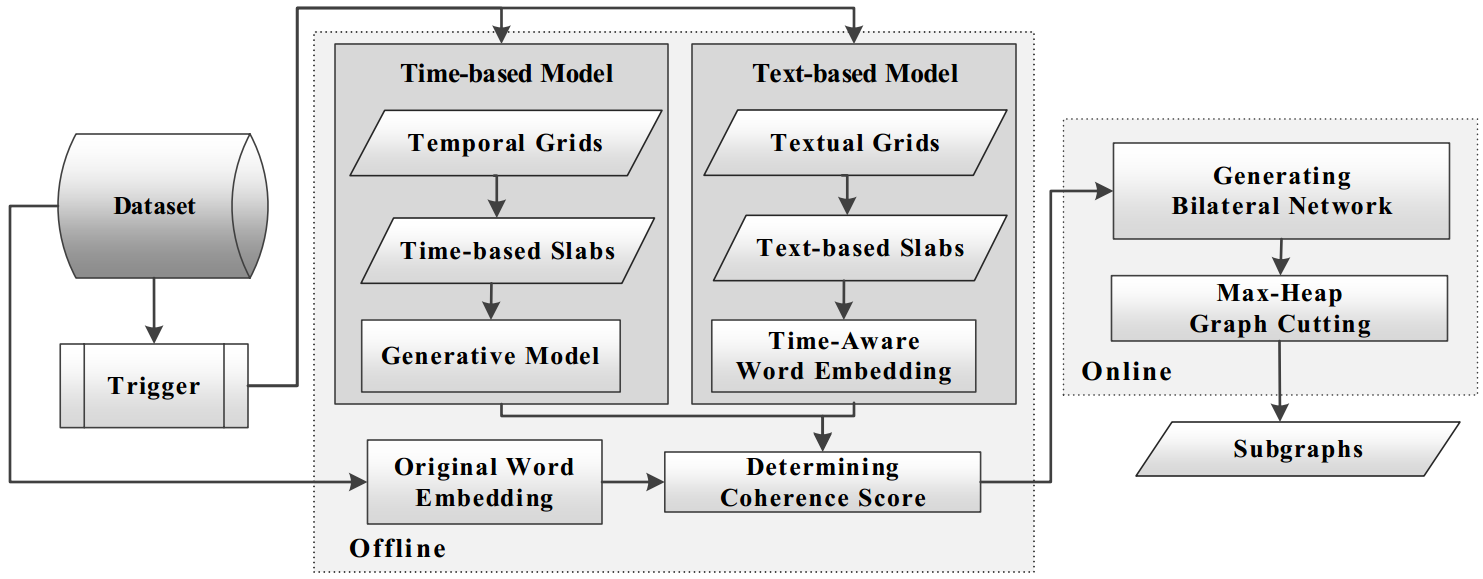}
					\vspace{-7mm}
					\caption{Framework}
					\vspace{-4mm}
					\label{fig:Framework}
				\end{figure}

		\vspace{-1mm}
		Figure \ref{fig:Framework} illustrates our proposed unified framework that can exploit the subgraphs, with highly correlated nodes, from propagation networks. 
		Through \underline{offline} phase, both the \textit{temporal} and \textit{textual} information of the alarms is used to retrieve the multi-aspect \textit{grids} that reflect how similar each pair of temporal
		splits are. Subsequently, we utilize \textit{hierarchical clustering} to combine similar intervals and form unified and then multi-aspect temporal \textit{slabs}.
		\vspace{-4mm}
		\vspace{-1mm}	
		In order to compute edge weights between nodes,	On the one hand, we feed the \textit{time-based slabs} to the \textit{multifacet generative model} to compute the time-only correlations, and on the other hand, the \textit{text-based temporal slabs} are consumed by the \textit{time-aware word embedding} to address the challenge that the co-occurrence patterns of the word pairs may differ in various temporal aspects. Moreover, the \textit{Original Word-Embedding} module takes the global co-occurrence matrix to retrieve non-temporal vector representation of the words in alarm contents. Nevertheless, the final \textit{coherence score}  can merge the edge weights computed by various methods.\\
		In the \underline{online} phase, the coherence score is used to merge the edge weights - computed by various models - to form a \textit{single-edged undirected graph}, named as \textit{Bilateral Network}. Finally, we employ the \textit{Max-Heap Graph cutting algorithm} to leverage the subgraphs with highly correlated nodes.
		\vspace{-4mm}
		\section{Methodology}
		\label{Methodology}
\begin{figure}[H]
										\centering
										\includegraphics[width=0.30\textwidth]{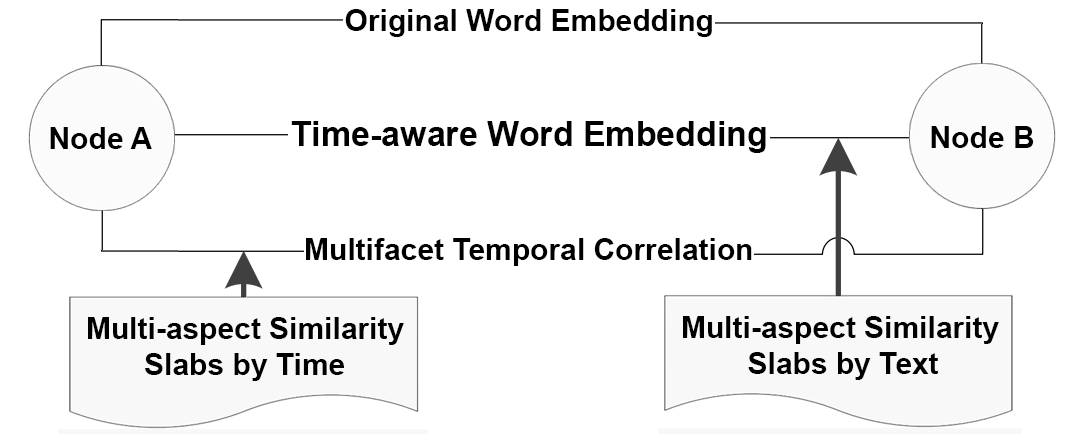}
										\vspace{-2mm}
										\caption{Inner Nodes correlations}
										\label{fig:InnerCorrelation}
										\vspace{-6mm}
									\end{figure}

		\vspace{-1mm}		
		In general, exploiting of the subgraphs with the highest correlated nodes primarily requires a series of \textit{isomorphic tests} to avoid double entries in the final list of subgraphs. Thus, \textit{canonical labeling} uniquely distinguishes every subgraph with a vector of digits. Accordingly, where a pair of subgraphs is canonically labeled the same, they can be considered \textit{isomorphic}. However, canonical labeling is computationally complex and declared as an \textit{NP-complete} problem.
		Therefore, in this paper, we alternatively propose a diligent algorithm, which initially transfers the problem of frequent subgraph mining to the computation of edge-weights. Conclusively, we devise a straightforward graph-cut algorithm to effectively leverage the communities of meticulously correlated nodes. Nevertheless, since the likelihood for every vertex in propagation graph is assigned by one, as Eq. \ref{Step_1} shows, the probability for each node $n_i$ to spread a contagion to another node $n_j$ corresponds to their joint probability.
		\vspace{-3mm}
		\begin{equation}
		\small
		\label{Step_1}
		Pr(n_j|n_i) \propto Pr(n_i,n_j)		
		\vspace{-2mm}
		\end{equation}
		We can collectively compute the edge weight between each pair of vertices (Fig. \ref{fig:InnerCorrelation}) using three algorithms, namely \textit{Original Word Embedding} (Sec. \ref{original-word-embedding}), \textit{Time-aware Word Embedding} (Sec. \ref{time-aware-word-embedding}), and \textit{Multifacet Generative Model} (Sec. \ref{Multifacet-Generative-Model}).
									\vspace{-3mm}
        \vspace{-3mm}
		\subsection{Offline Processing}				
		\vspace{-1mm}
		\subsubsection{Constructing Similarity Grids}
		\label{Constructing-Similarity-Grids}
		\vspace{-1mm}
		Similarity grids consider the \textit{Temporal} and \textit{Textual} propagation rules in each temporal dimension to discover correlation rates between pair of splits. We employ two real-life datasets (Section \ref{Data}) for empirical studies.		
		\vspace{-5mm}
		\begin{figure}[!htp]
			\centering
			\minipage{0.24\textwidth}
			\centering
			\includegraphics[width=\linewidth]{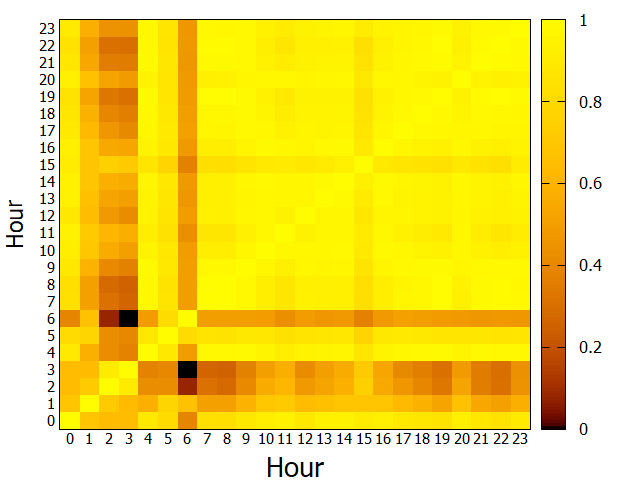} 
			\small (a) Hour ($z^h$)
			\endminipage\hfill	
			\minipage{0.24\textwidth}
			\centering
			\includegraphics[width=\linewidth]{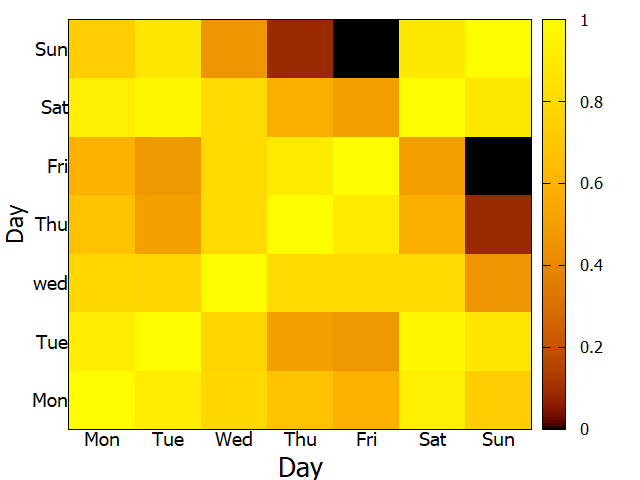} 
			\small (b) Day ($z^d$)
			\endminipage\hfill
			\minipage{0.24\textwidth}
			\centering
			\includegraphics[width=\linewidth]{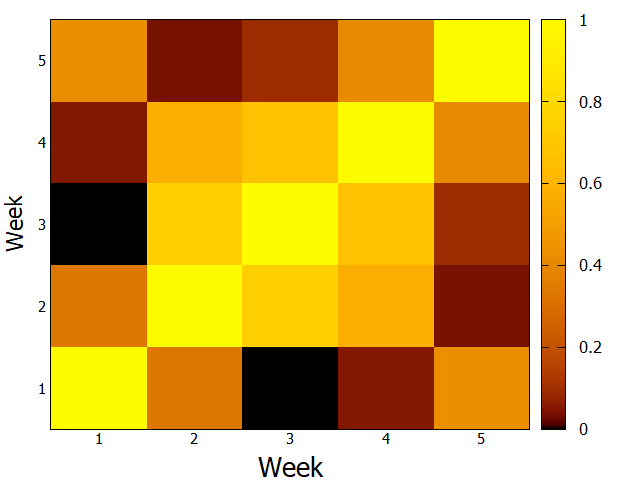} 
			\small (c) Week ($z^w$)
			\endminipage\hfill
			\minipage{0.24\textwidth}
			\centering
			\includegraphics[width=\linewidth]{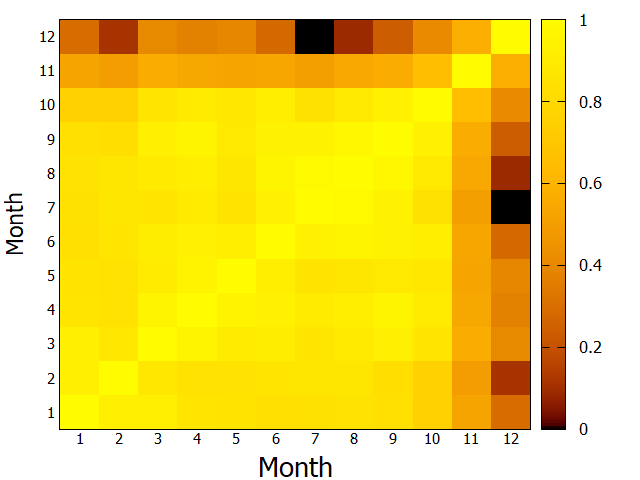} 
			\small (d) Month ($z^m$)
			\endminipage				
			\vspace{-2mm}
			\caption{Similarity Grid - Textual Data - ST-0}
			\vspace{-5mm}
			\label{fig:Textual_Similarity_Slots_ST0}			
		\end{figure}
		\begin{figure}[!htbp]
			\centering
			\minipage{0.24\textwidth}
			\centering
			\includegraphics[width=\linewidth]{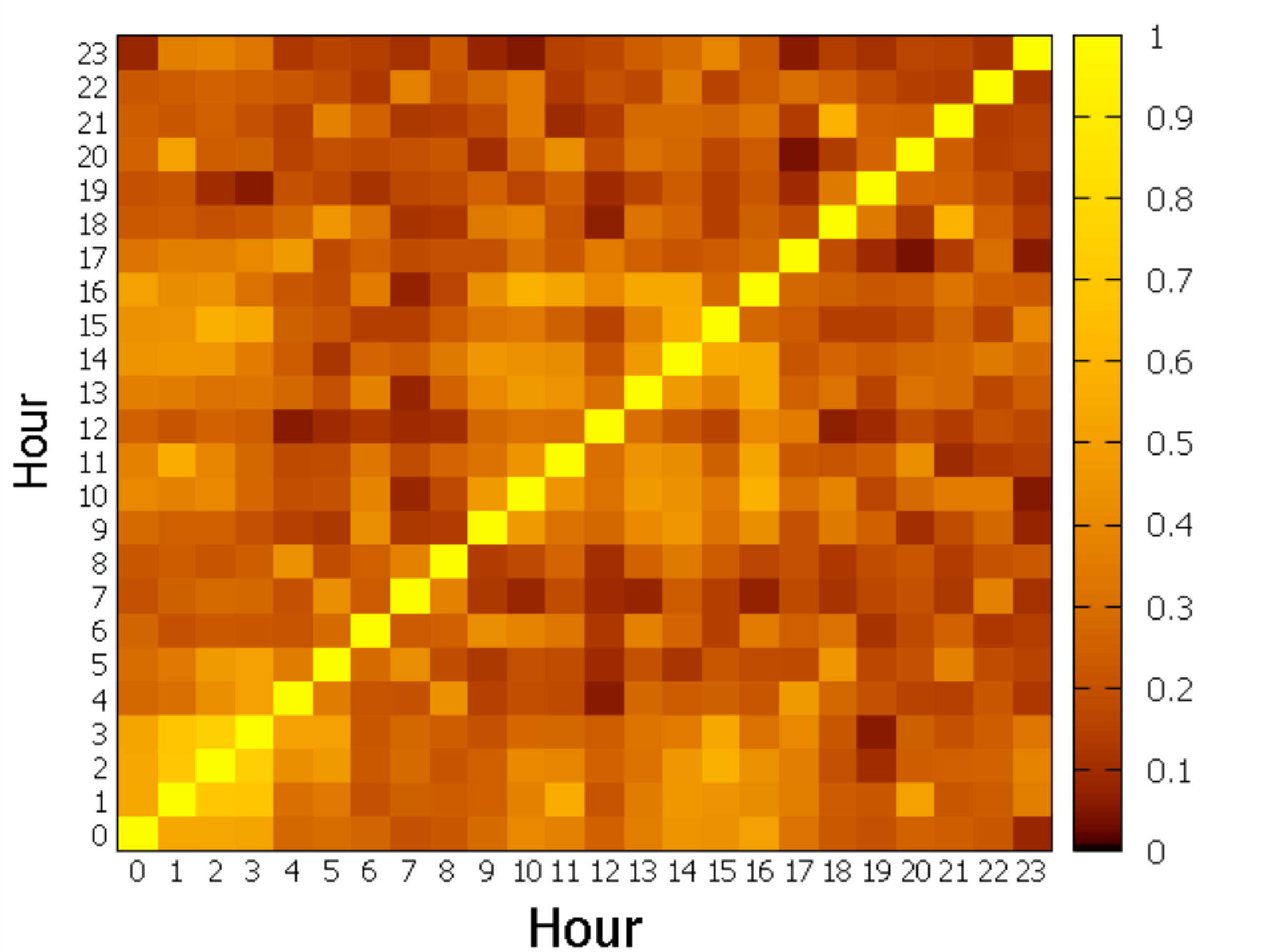} 
			\small (a) Hour ($z^h$)
			\endminipage\hfill	
			\minipage{0.24\textwidth}
			\centering
			\includegraphics[width=\linewidth]{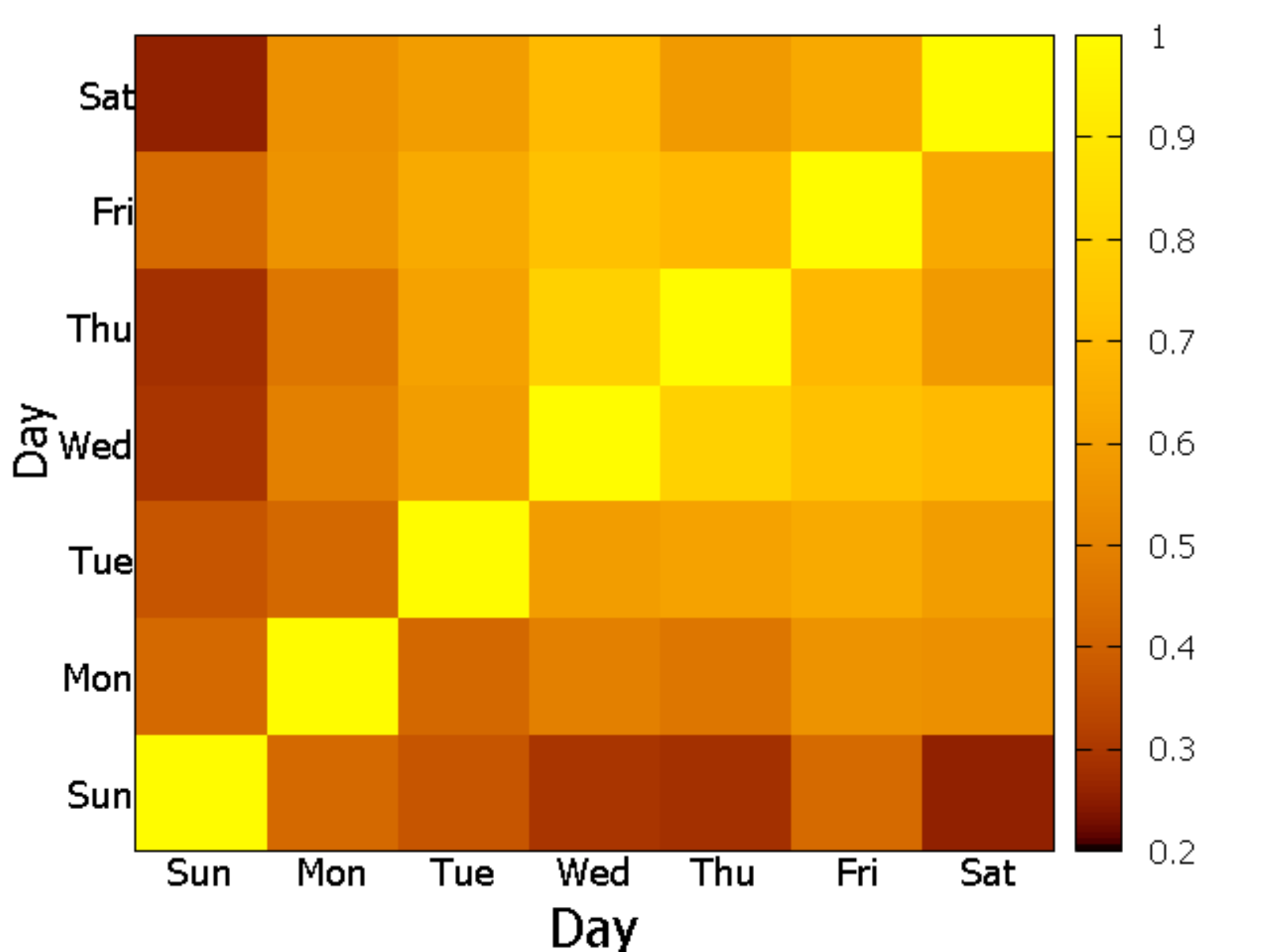} 
			\small (b) Day ($z^d$)
			\endminipage\hfill
			\minipage{0.24\textwidth}
			\centering
			\includegraphics[width=\linewidth]{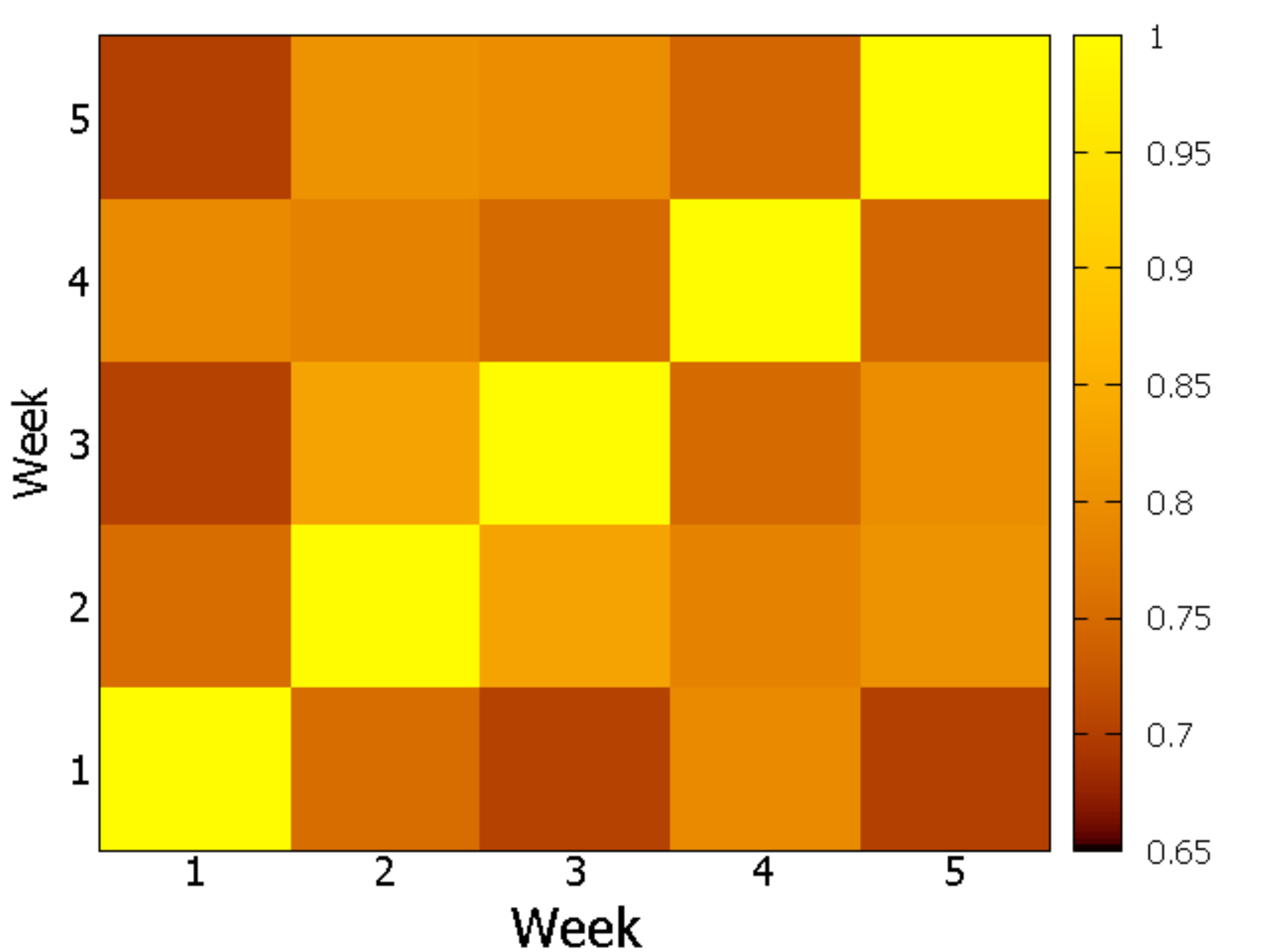} 
			\small (c) Week ($z^w$)
			\endminipage\hfill
			\minipage{0.24\textwidth}
			\centering
			\includegraphics[width=\linewidth]{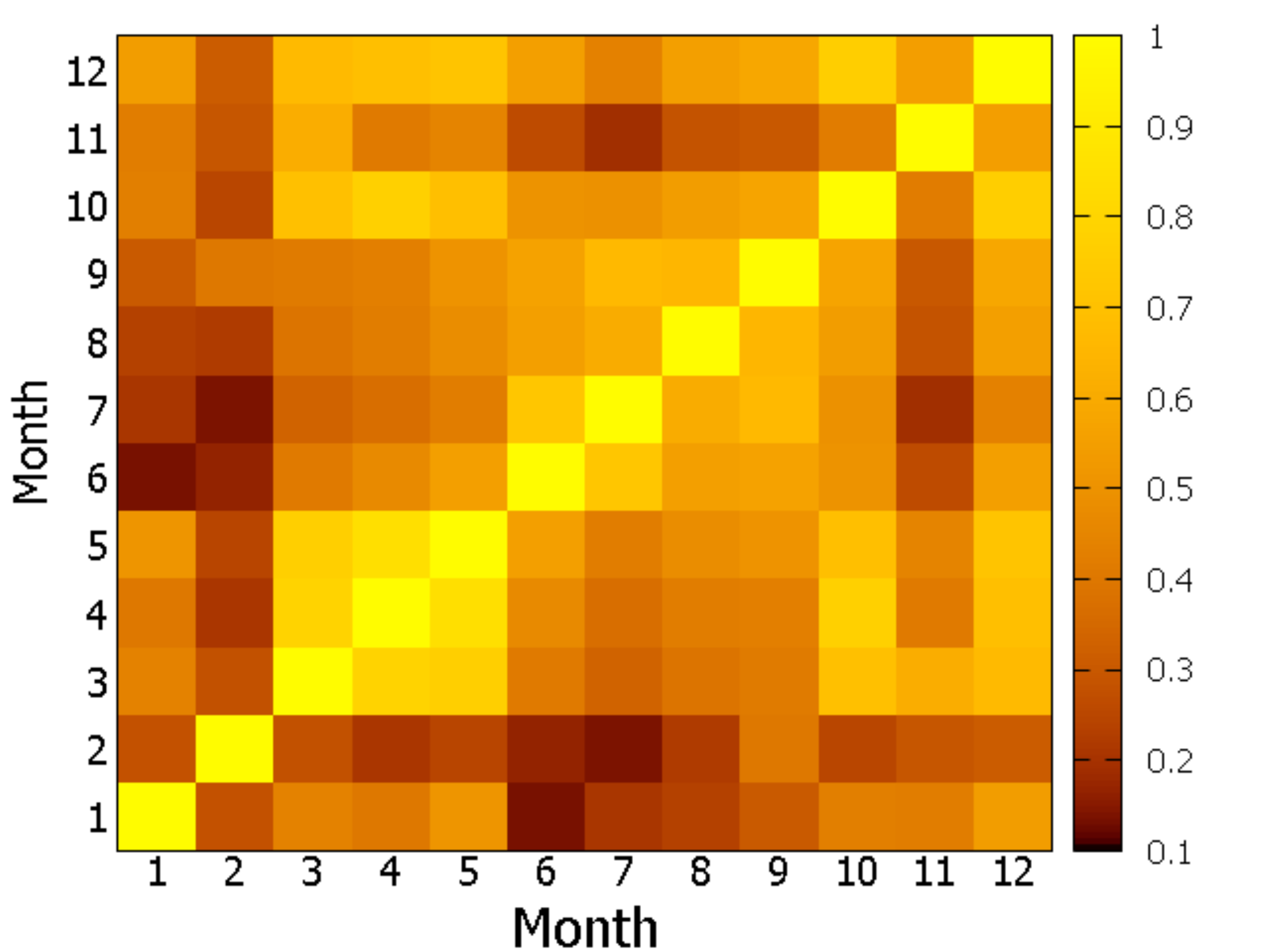} 
			\small (d) Month ($z^m$)
			\endminipage				
			\vspace{-2mm}			
			\caption{Similarity Grid - Temporal Data - ST-0}
			\label{fig:Similarity_Slots_ST0}
			\vspace{-4mm}
		\end{figure}
		\textbf{Textual Grid}: In order to construct the similarity grid using textual contents, we firstly aggregate the textual contents of all the nodes during each temporal split, say \textit{5 am} in \textit{hour} dimension. Hence, we associate an array of text with each latent facet. Note that, each cell in $\mathbb{NTA}$ cube additionally adds the textual contents of the alarms in the pertinent split. Subsequently, we employ a modified version of \textit{TF-IDF}, to signify the weight of each word in every latent facet:
		\vspace{-2mm}
		\begin{equation} 
			\small
			\label{eq:TFIDF}
			\mathbf{\hat{w}}(t_i,S_k^l)=\frac{f(t_i,S_k^l)}{Max_{(t\in S^l)}\{f(t,S_k^l)\}} \times Log \frac{N}{N(t_i)}
			\vspace{-2mm}
		\end{equation}
		As verbalized in Eq. \ref{eq:TFIDF}, $S_k^l$ constitutes the string contents of split $k$ in latent facet $l$. Also, $\frac{f(t_i,S_k^l)}{Max_{(t\in S^l)}\{f(t,S_k^l)\}}$ estimates the normalized term frequency in the split. Here, $N$ designates the total number of the splits and $N(t_i)$ is the number of splits comprising the term $t_i$. Given the computed list of normalized weights, every split can be presented by a fixed-sized vector $\vec{S}_k^l$, wherein the entries show the weights of the terms. The split vectors are then employed by the \textit{Cosine} metric to obtain the similarity between splits.\\
		Figure \ref{fig:Textual_Similarity_Slots_ST0} illustrates the textual similarity grids. Since the contrast between the textual contents of the splits in hour and month dimensions are low, our model grants more significance to other facets (day and week) for ST-0 dataset.\\		
		\noindent \textbf{Temporal Grid}: 
		In order to generate the temporal similarity grid for each latent facet (e.g. Hour), we find the nodes that are infected in each pair of the splits. Take the hour facet, where the same alarm type happens in both hours, we can claim an evidence that two hours are partially similar. In contrary, if the number of mutual alarm types between a pair of hours is zero, one can conclude that the two hours are mutually exclusive or at least we cannot decide whether we can cluster them together or not. Nevertheless, we use the \textit{linear correlation measure} to compute the final similarity weight between each split pairs in each dimension. Note that the resulting unifacet temporal slabs are affected by the sampling model based on which the similarity evidences are collected. Accordingly, we employ the \textit{stratified sampling model} \cite{Yan2014}. Aiming to achieve $q$ similarity evidences between split pairs, we process $p$\% of the nodes through a \textit{replacing iterative process}. Furthermore, we utilize the \textit{None-Negative Matrix Factorization (NMF)}  to compensate missing entries in each similarity grid. Figure \ref{fig:Similarity_Slots_ST0} demonstrates the time-only similarity grids for datasets ST-0. While neighboring intervals show identical similarity weights in recommendation systems \cite{hosseini2017leveraging}\cite{Hosseini2016}, we interestingly observe in Figure \ref{fig:Similarity_Slots_ST0} that the similarity grids in diffusion networks differ and consequently need to get updated via the \textit{Trigger} frequently.
				
		\vspace{-3mm}
		\subsubsection{Acquiring Temporal Slabs}		
		\label{Acquiring-Temporal-Slabs}
		\vspace{-1mm}
		Given the similarity grids, computed using temporal and textual evidences, and also the density of dataset, we employ the bottom-up \textit{Hierarchical Agglomerative Clustering} (\textit{HAC} via \textit{complete linkage}) to retrieve temporal similarity slabs. In this way, we address Problem \ref{extracting-multifacet-temporal-slabs} and extract multifacet temporal slabs. The number of dimensions, e.g. quadrilateral facets ($z^h,z^d,z^w,z^m$) can be affirmed based on the sparsity of dataset. Nevertheless, the clustering \textit{threshold} is utterly important as it can incorporate irrelevant temporal splits into the same slab or toss pertinent splits to separate clusters. To address this challenge, we merge the splits in each dimension based on various thresholds that can construct two series of slabs, i.e. \textit{small} and \textit{big clusters}. Subsequently, either of big or small clusters, that can ensure the best effectiveness in subgraph mining (Section \ref{benchmark}), will be consumed by the inference models (Section \ref{Impact-of-zeta-on-effectiveness-of-Big-and-Small-Clusters}).\\
		\vspace{-8mm}
		\begin{figure}[H]
			\centering
			\minipage{0.10\textwidth}
			\centering
			\includegraphics[width=\linewidth]{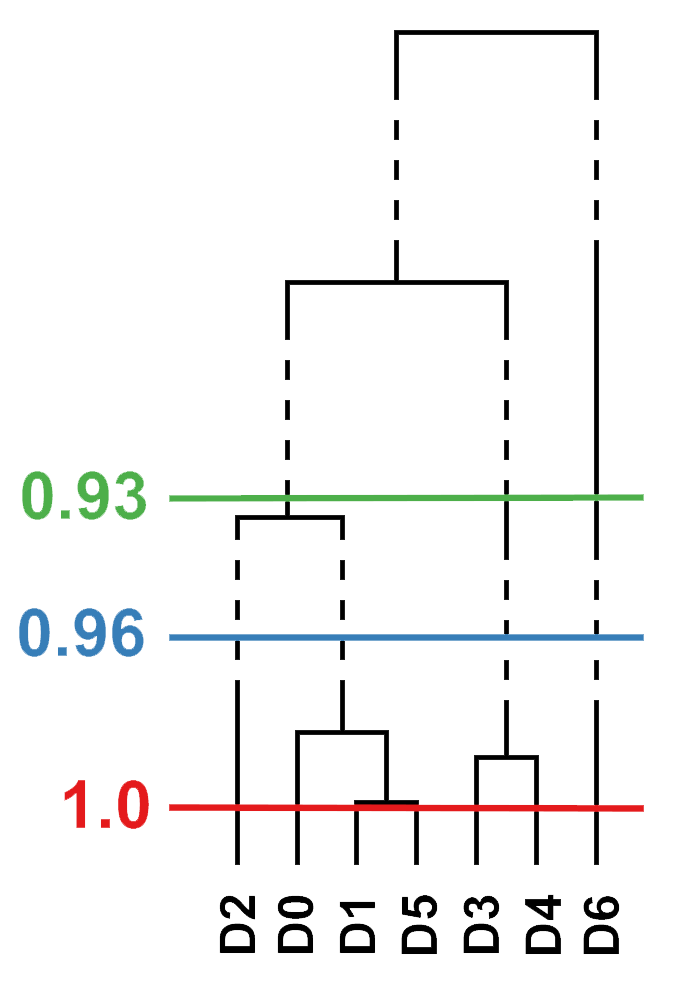} 
			\small (a) ST-0 Day HAC ($z^h$)
			\endminipage\hfill	
			\minipage{0.10\textwidth}
			\centering
			\includegraphics[width=\linewidth]{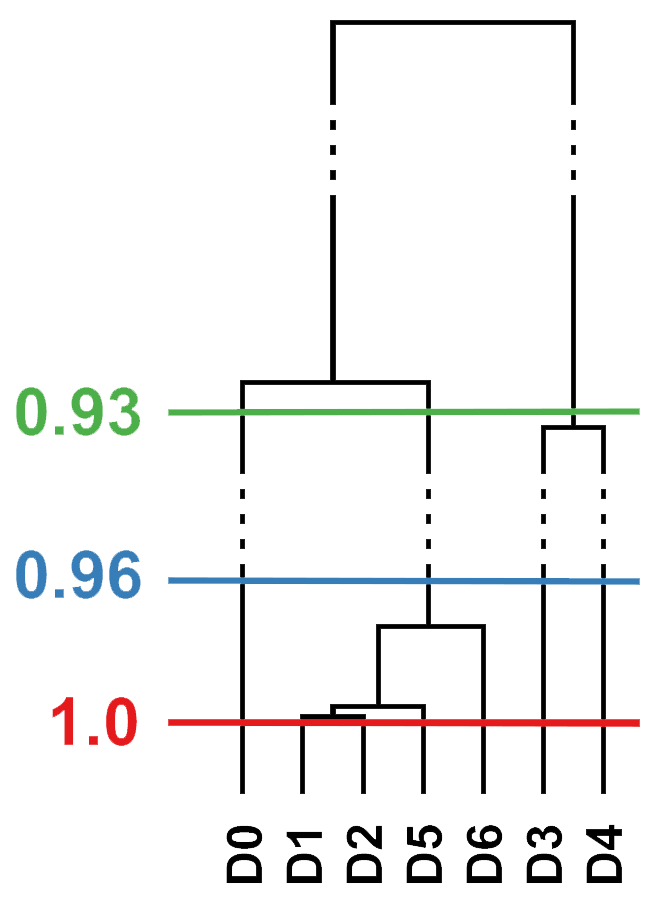} 
			\small (b) ST-1 Day HAC ($z^d$)
			\endminipage\hfill
			\minipage{0.10\textwidth}
			\centering
			\includegraphics[width=\linewidth]{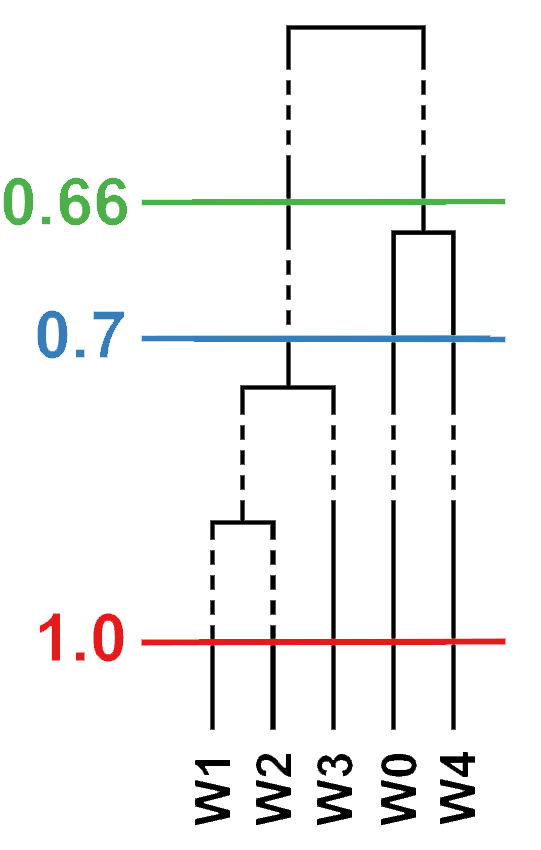} 
			\small (c) ST-0 Week HAC ($z^w$)
			\endminipage\hfill
			\minipage{0.10\textwidth}
			\centering
			\includegraphics[width=\linewidth]{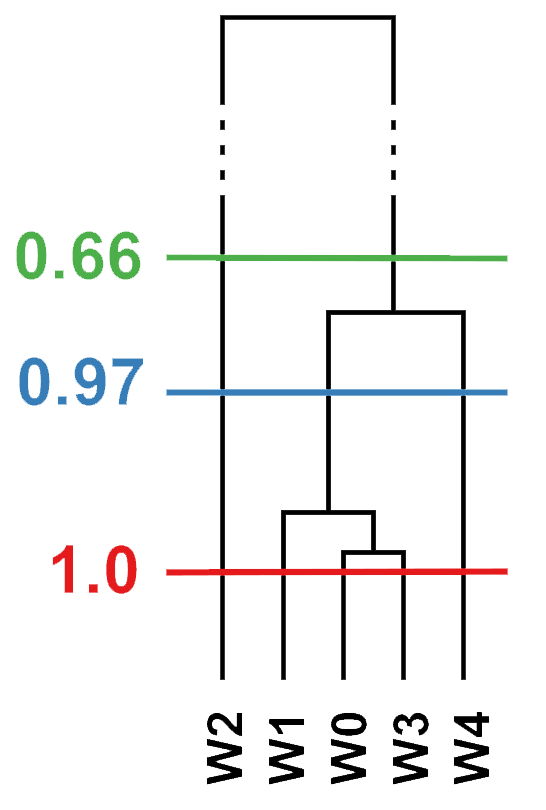} 
			\small (d) ST-1 Week HAC ($z^m$)
			\endminipage
			\vspace{-3mm}				
			\caption{Sample dendrograms - Big versus Small clusters}
			\label{fig:Similarity_Slots_ST1}
		\end{figure}
		\vspace{-4mm}
		Considering two daily and weekly facets, Fig. \ref{fig:Similarity_Slots_ST1} illustrates distinctive thresholds which will generate either of small (middle line) or big clusters (upper line). Accordingly, Table \ref{simialr_textual_slabs} reports generated clusters in four-fold temporal dimensions. As empirically  witnessed in Section \ref{Impact-of-zeta-on-effectiveness-of-Big-and-Small-Clusters}, compared with small clusters, the big clusters can better exploit the tightly connected subgraphs in both datasets.
		\vspace{-2mm}
		\begin{table}[H]
			\centering
			\caption{Small and Big Clusters(Textual evidences)}	
			\vspace{-3mm}
			\label{simialr_textual_slabs}
			\begin{tabular}{c|c|c|c|}
				\cline{2-4}
				& Station               & Big Clusters          & Small Clusters            \\ \hline
				\multicolumn{1}{|c|}{}                                                                      &                       & \{1\}\{2,3\}\{6\}\{0, & \{1\}\{2,3\}\{5\}\{6\}    \\
				\multicolumn{1}{|c|}{}                                                                      & \multirow{2}{*}{ST-0} & 11,15\}\{4,5,7,8,9,   & \{0,11,15\}\{4,7,8,9,     \\
				\multicolumn{1}{|l|}{}                                                                      &                       & 10,12,13,14,16,17,    & 18,19,21,22,23\}\{10,     \\
				\multicolumn{1}{|c|}{\multirow{2}{*}{\begin{tabular}[c]{@{}c@{}}Hour\\ Slabs\end{tabular}}}  &                       & 18,19,20,21,22,23\}   & 12,13,14,16,17,20\}       \\ \cline{2-4} 
				\multicolumn{1}{|c|}{}                                                                      &                       & \{0\}\{1,4,5\}\{2,3,  & \{0\}\{1,4\}\{5\}\{6,23\} \\
				\multicolumn{1}{|l|}{}                                                                      & \multirow{2}{*}{ST-1} & 6,7,8,9,10,11,12,     & \{2,3,7,8,9,10,11,        \\
				\multicolumn{1}{|c|}{}                                                                      &                       & 13,14,15,16,17,18,    & 12,13,14,15,16,           \\
				\multicolumn{1}{|c|}{}                                                                      &                       & 19,20,21,22,23\}      & 17,18,19,20,21,22\}       \\ \hline
				\multicolumn{1}{|c|}{}                                                                      & \multirow{2}{*}{ST-0} & \{Mon,Tue,Wed,Sat\}   & \{Mon,Tue,Sat\}\{Wed\}    \\
				\multicolumn{1}{|c|}{\multirow{2}{*}{\begin{tabular}[c]{@{}c@{}}Day\\ Slabs\end{tabular}}}   &                       & \{Thu,Fri\}\{Sun\}    & \{Thu,Fri\}\{Sun\}        \\ \cline{2-4} 
				\multicolumn{1}{|c|}{}                                                                      & \multirow{2}{*}{ST-1} & \{Tue,Wed,Sat,Sun\}   & \{Tue,Wed,Sat,Sun\}       \\
				\multicolumn{1}{|c|}{}                                                                      &                       & \{Mon\}\{Thu,Fri\}    & \{Mon\}\{Thu\}\{Fri\}     \\ \hline
				\multicolumn{1}{|c|}{\multirow{2}{*}{\begin{tabular}[c]{@{}c@{}}Week\\ Slabs\end{tabular}}}  & ST-0                  & \{1,2,3\}\{0,4\}      & \{0\}\{1,2,3\}\{4\}       \\ \cline{2-4} 
				\multicolumn{1}{|c|}{}                                                                      & ST-1                  & \{2\}\{0,1,3,4\}      & \{0,1,3\}\{2\}\{4\}       \\ \hline
				\multicolumn{1}{|c|}{}                                                                      & \multirow{2}{*}{ST-0} & \{0,1,6,7\}\{10,11\}  & \{0,1\}\{6,7\}\{10,11\}   \\
				\multicolumn{1}{|c|}{\multirow{2}{*}{\begin{tabular}[c]{@{}c@{}}Month\\ Slabs\end{tabular}}} &                       & \{2,3,4,5,8,9\}       & \{2,3,4,5,8,9\}           \\ \cline{2-4} 
				\multicolumn{1}{|c|}{}                                                                      & \multirow{2}{*}{ST-1} & \{0\}\{11\}\{1,2,3,   & \{0\}\{11\}\{1,2,10\}     \\
				\multicolumn{1}{|c|}{}                                                                      &                       & 4,5,6,7,8,9,10\}      & \{3,4,5,6,7,8,9\}         \\ \hline
			\end{tabular}
			\vspace{-4mm}
		\end{table}				
		\textbf{Trigger}. Owing to diffusions in propagation networks, the similarity values between temporal splits may change. The \textit{Trigger} can thus estimate under what intervals we must update the similarity grids. This can negatively affect leveraged slabs.
		Hence, we propose an effective algorithm based on \textit{divide and expansion} to approximate trigger interims.
		\vspace{-3mm}
		\begin{algorithm}[H]
			\caption{Trigger Interval Estimation}
			\label{triggering}
			\textbf{Input:} $\epsilon$, $\mathbb{D}_{e}$, $\delta$\\
			\textbf{Output:} $\Delta$ \par
			\begin{algorithmic}[1]
				\STATE $\mathbb{H}=$ Split($\mathbb{D}_{e}, 2$)
				\STATE $\tilde{G} =$ $Sub_g$($\mathbb{H}[0]$)
				\STATE $\kappa^0 = \boldsymbol{{E}}_v(\tilde{G}) $
				\STATE $q_s = \mathbb{H}[0]$
				\FOR{each $i$ in Range(1, $|\mathbb{H}[1]|/\delta$)}
				\STATE $q_s= q_s +$ Subset($\mathbb{H}[1], i$)
				\STATE $\tilde{q_s}=$ $Sub_g$($q_s$)
				\STATE $\kappa^i = \boldsymbol{{E}}_v(\tilde{q_s}) $
				\IF{$\kappa^0 - \kappa^i>=\epsilon$}
				\STATE Return\;$i*\delta$
				\ENDIF
				\ENDFOR
				\STATE Return\;$|\mathbb{H}[1]|$
			\end{algorithmic}
		\end{algorithm}
		\vspace{-4mm}
		As demonstrated in Algorithm \ref{triggering}, we first take the whole or a portion of the dataset (denoted by $\mathbb{D}_{e}$) and \textit{divide} it into two equal subsets ($\mathbb{H}$=$\{\mathbb{H}[0],\mathbb{H}[1]\}$). The algorithm is capable to consume a subset of the dataset to accomplish the task quicker.
		To start, we use the first subset $\mathbb{H}[0]$ to construct the similarity grids and acquire the temporal slabs, based on which we can employ any model $Sub_g$ to exploit the set of subgraphs $\tilde{G}$. The evaluation function $\boldsymbol{{E}}_v$ can afterward measure the effectiveness $\kappa^0$ using benchmark (Section \ref{benchmark}). While consuming the same input grids, we then incrementally \textit{expand} the first half $\mathbb{H}[0]$ by the size of $\delta$, which is the portion from the second half $\mathbb{H}[1]$. After each expansion, we reevaluate the effectiveness $\kappa^i$ in the current iteration $i$.
		In summary, the deficiency rate ($\kappa^0-\kappa^i$) elucidates how the unchanged similarity grids can escalate errors when the data size grows. Consequently, when the error size is greater than $\epsilon$, the algorithm will return the final \textit{tolerance threshold} which is the expansion size of the current iteration $i*\delta$. Therefore, when the size of the data grows by the tolerance threshold, it will enforce the rebuilding of the input grids. If the error rate never exceeds the $\epsilon$, the returned value $|\mathbb{H}[1]|$ will indicate that either the initial portion, i.e. $\mathbb{D}_{e}$, is chosen incorrectly or the similarity changes are negligible.
		\vspace{-2mm}
		\subsubsection{Original Word Embedding (\textit{OWE})}
		\vspace{-1mm}
		\label{original-word-embedding}
	When the semantic intersection between $O_u$ and $O_v$ as the respective textual contents of the node pair ($u,v$) is insignificant, the state-of-the-art text mining models, e.g. \cite{nguyen2015improving}, cannot perceive intra-nodes correlations. However, the semantic vector space models \cite{Mikolov2013a}\cite{Deerwester1990} adopt the dataset-wide knowledge of word co-occurrences to retrieve the vector representation of each word and construct the list $\vec{w}_i^o$ of similar words to $w_i$ in descending order. Accordingly, we can produce a newly formed \textit{encyclopedic semantic representation} $O^\prime_u$ for the each vertex $u$ through replacing any word $w_i \in O_u$ by the top $\zeta$ most similar words from $\vec{w}_i^o$. To this end, we utilize \textit{Original Word Embedding} (OWE) \cite{Pennington2014}, which is a weighted least square regression model (Eq. \ref{J1-loss-function}).
		\vspace{-2mm}
		\begin{equation}
			J_{owe}=\sum_{\forall w_i,w_j \in \mathbb{V}} F(X_{ij})\times((\vec{w}_i^m)^{T}(\vec{w}_j^c)+b_i^m+{b}_j^c-LogX_{ij})^2
			\label{J1-loss-function}
		\end{equation}		
		Here, $X$ is the co-occurrence matrix and every entry $X_{ij}$ counts the number of times any word $w_j$ occurs in the context of $w_i$. Also, $\vec{w}_i^m$ and $\vec{w}_j^c$ outline the main ($w_i$ in the center of sliding-window) and context ($w_i$ co-occurs with any other word in the center of window) vectors.
		Moreover, the cutoff function $F(X_{ij})$ avoids overweighting of the rare or frequent co-occurrence patterns.
		Finally, each word $w_i$ will be assigned by the median of main $\vec{w}_m^i$ and context $\vec{w}_c^i$ vectors \cite{dumais2004latent}, denoted by $\vec{w_i}$.
		\vspace{-2mm}
		\begin{equation}
			(\vec{w}_i^m)^{T}(\vec{w}_j^c)+b_i^m+{b}_j^c=LogX_{ij}
			\vspace{-1mm}
			\label{Equality_Equ_GloVe}
		\end{equation}
		While randomly initialized $\vec{w}_i^m$ and $\vec{w}_j^c$ vectors converge toward $Log(X_{ij})$, the bias parameters of $b_i^m$ and ${b}_j^c$ ensure the symmetry. Our framework utilizes OWE module to infer non-temporal co-occurrence patterns in word embedding.
		\vspace{-2mm}
		\subsubsection{Time-aware Word Embedding (\textit{TWE})}
		\label{time-aware-word-embedding}
\begin{figure} [H]
					\centering
					\includegraphics[width=0.49\textwidth]{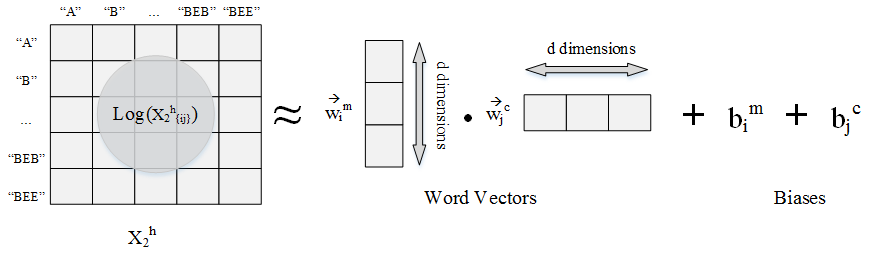}
					\vspace{-9mm}
					\caption{Time-aware word embedding}			
					\label{fig:WordEmbedding}
					\vspace{-4mm}
				\end{figure}

		\vspace{-1mm}
		As explained in Section \ref{sec:introduction}, the word-word co-occurrence patterns are temporally skewed and change in various latent facets. However, current word embedding approaches \cite{dumais2004latent}\cite{Mikolov2013a}\cite{Pennington2014}, including \textit{OWE}, neglect such temporal dynamics. Thus, we devise a novel time-aware embedding module to track co-occurrence alterations and further infer the \textit{multi-aspect temporal-semantical balance} between the nodes. We first despise the biases in OWE \cite{Pennington2014} and changes Eq. \ref{Equality_Equ_GloVe} to Eq. \ref{Equality_Equ_GloVe_01}.
		\vspace{-2mm}
		\begin{equation}
			(\vec{w}_i^m)^{T}(\vec{w}_j^c) \propto LogX_{ij}
			\label{Equality_Equ_GloVe_01}
			\vspace{-2mm}
		\end{equation}
				Also, Eq.\ref{timeaware_01} forms the \textit{single slab} time-aware version of Eq. \ref{Equality_Equ_GloVe_01}.\\
						\vspace{-2mm}
						\begin{equation}
						\begin{split}
						{\forall w_i,w_j \in \mathbb{V}_k^l}, (\vec{w_i^m})^T(\vec{w_j^c})\propto \mathcal{F}(w_i,w_j,z_k^l)\times \rho \\ \propto Log(X^l_{k\{i,j\}})
						\end{split}
						\vspace{-3mm}
						\label{timeaware_01}
						\end{equation}
				From one side, the \textit{slab-based coefficient function} $\mathcal{F}(w_i,w_j,z_k^l)$ corresponds to the inner product of the main and context vectors, and from the other side, it represents the log-based tally for word pair ($w_i,w_j$) co-occurrence in the same slab $z_k^l$. For instance, as depicted in Fig. \ref{fig:WordEmbedding}, $z_2^h$ is the second slab in the hour facet. Nevertheless, for ease of exposition, we will explain the types of coefficient functions later this Section.		
				\vspace{-4mm}
								\vspace{-8mm}
								\begin{figure} [H]	
									\centering
									\includegraphics[width=0.30\textwidth]{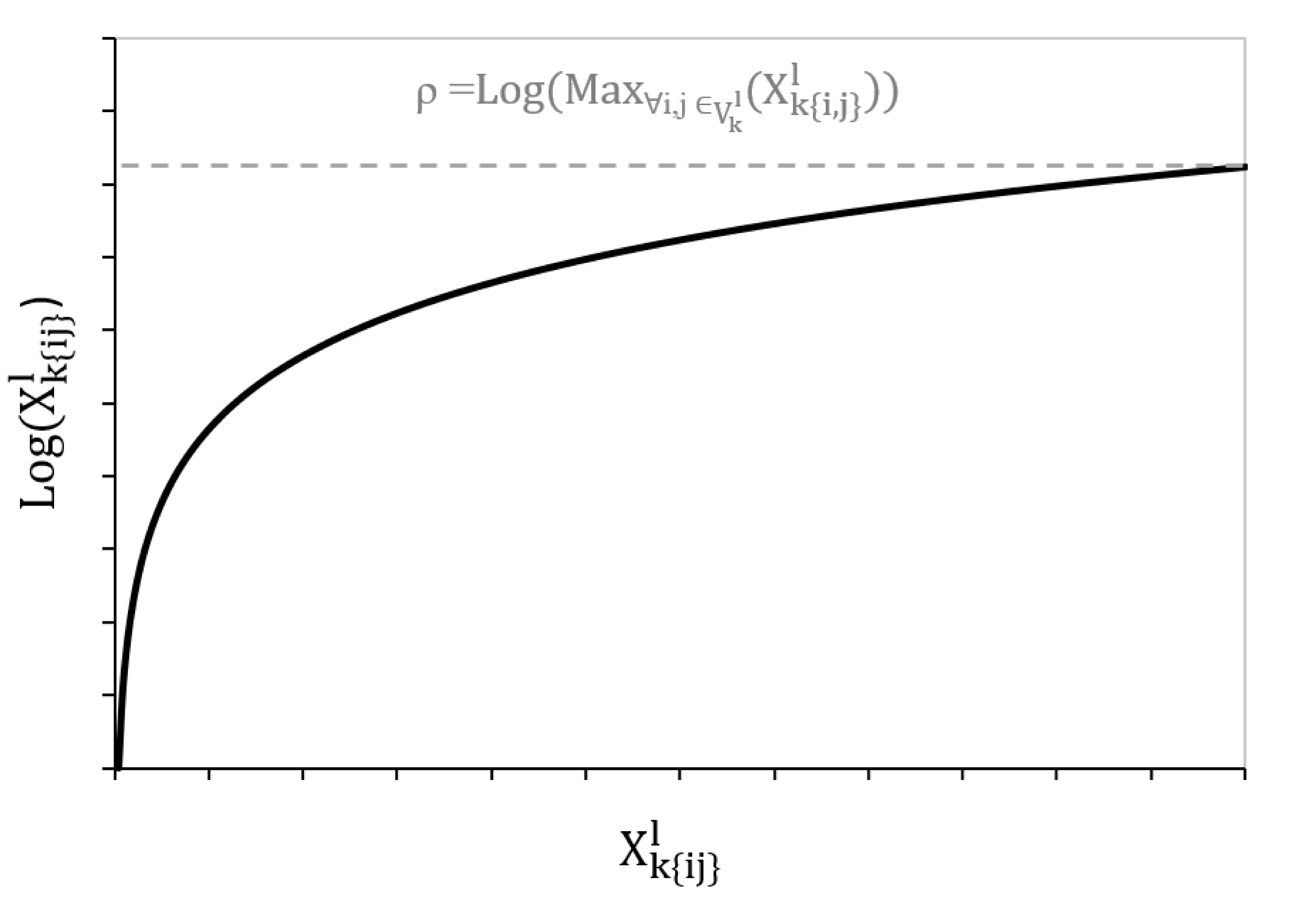}
									\vspace{-3mm}
									\caption{Log chart}
									\label{fig:LogChart}
								\end{figure}
								\vspace{-4mm}
								
				As Fig. \ref{fig:LogChart} shows, the upper bound parameter $\rho$ imposes that both sides of Eq. \ref{timeaware_02} will get bounded from above ($U_b$).
		Where the maximum value for the slab-based coefficient is 1, the value for $\rho$ can be affirmed by Eq. \ref{timeaware_03}.
		\vspace{-2mm}
				\begin{equation}
				U_b(\mathcal{F}(w_i,w_j,z_k^l) \times \rho)^\uparrow = U_b(Log(X^l_{k\{i,j\}}))^\uparrow
				\label{timeaware_02}			
				\vspace{-2mm}
				\end{equation}
		\vspace{-3mm}
		\begin{equation}
			\rho = Log(Max_	{\forall w_i,w_j \in \mathbb{V}_k^l}(X^l_{k\{i,j\}}))
			\label{timeaware_03}
		\end{equation}
		We can then attain Eq. \ref{timeaware_05} via substituting of Eq. \ref{timeaware_03} in Eq. \ref{timeaware_01}.
		\vspace{-2mm}
		\begin{equation}
			\begin{split}
					{\forall w_i,w_j \in \mathbb{V}_k^l}, (\vec{w_i^m})^T(\vec{w_j^c})\propto \mathcal{F}(w_i,w_j,z_k^l)\\ \times Log(Max_{\forall w_i,w_j \in \mathbb{V}_k^l}(X^l_{k\{i,j\}}))
			\end{split}
			\vspace{-8mm}
			\label{timeaware_05}			
		\end{equation}		
		The symmetry in Eq. \ref{timeaware_05} is preserved via the log-based value of $Max_{\forall w_i,w_j \in \mathbb{V}_k^l}(X^l_{k\{i,j\}})$ on the right side. Considering the independent appearance for $i$ and $j$, adding $b_i^m$ and $b_j^c$ to Eq. \ref{timeaware_05} can further maintain the symmetry in the final Eq. \ref{timeaware_06}.\\
		\vspace{-2mm}
		\begin{equation}
			\begin{split}
				{\forall w_i,w_j \in \mathbb{V}_k^l}, (\vec{w_i^m})^T(\vec{w_j^c}) + b_i^m + b_j^c = \mathcal{F}(w_i,w_j,z_k^l)\\
				\times Log(Max_{\forall w_i,w_j \in \mathbb{V}_k^l}(X^l_{k\{i,j\}})
			\end{split}
			\vspace{-2mm}
			\label{timeaware_06}
		\end{equation}		
		To continue, we clarify our solution for slab-based coefficient function. Our idea is to invoke two attributes for the slab-based coefficiency (i.e. $\mathcal{F}(w_i,w_j,z_k^l)$) to better infer correlation intensity between each pair of words ($w_i,w_j$).
		\vspace{-2mm}
		\begin{itemize}
			\item \textit{Contiguity} ($\mathcal{F}_{con}(w_i,w_j,z_k^l)$) explains how extended each pair of words co-occur together in each of the temporal slabs.
			\item \textit{Depth} ($\mathcal{F}_{dep}(w_i,w_j,z_k^l)$) infers how a pair of words co-occur in each slab when the impacts from parent temporal dimension(s) are inevitably considered.
		\end{itemize}
		\vspace{-1mm}
		Subsequently, we put forward a distinctive weighted least square regression model for either of contiguity (Eq. \ref{con_regression}) and depth (Eq. \ref{dep_regression}) attributes of the slab-based coefficient.\\
		\vspace{-2mm}
		\begin{equation}
		\begin{split}
		L_{\phi}(z_k^l)=\sum_{\forall w_i,w_j \in  \mathbb{V}_k^l}f(X_{k\{ij\}}^l)((\vec{w_i^m})^T(\vec{w_j^c})+b_i^m+b_j^c\\-\mathcal{F}_{con}(w_i,w_j,z_k^l)\times Log(Max(\forall X_{k\{ij\}}^l)))^2
		\end{split}
		\label{con_regression}
		\end{equation}
		\vspace{-1mm}
		\begin{equation}
		\begin{split}
		L_{\theta}(z_k^l)=\sum_{\forall w_i,w_j \in  \mathbb{V}_k^l}f(X_{k\{ij\}}^l)((\vec{w_i^m})^T(\vec{w_j^c})+b_i^m+b_j^c\\-\mathcal{F}_{dep}(w_i,w_j,z_k^l)\times Log(Max(\forall X_{k\{ij\}}^l)))^2
		\end{split}
		\label{dep_regression}
		\end{equation}
		\vspace{-1mm}
				\begin{equation}		
				F(x)=\begin{cases}
				(x/x_{max})^\alpha \quad \quad \quad \quad if \; x<x_{max}\\
				1 \quad \quad \quad \quad \quad \quad \quad \quad  otherwise.\\
				\end{cases}
				\label{class-function}
				\vspace{-2mm}
				\end{equation}
				Furthermore, as formalized in Eq. \ref{class-function}, we incorporate the weighting cut-off function $F(X_{k\{ij\}}^l)$ \cite{Pennington2014} into the regression model and restrict mutual frequencies to the current slab (i.e. slab $k$ in temporal dimension of $l$). The class function precludes overweighting of the exceptional or periodic co-occurrence patterns. Moreover, we apply similar parameter settings of \cite{Pennington2014} and \cite{Mikolov2013a}.
				
		Finally, we propose two loss functions of $J_{twe}^c$ and $J_{twe}^c$ that respectively address the problem of time-aware word embedding through contiguity and the depth attributes. As observed in Eq. \ref{loss-functions-final}, both regression models iterates through the slabs and latent temporal facets.
\vspace{-4mm}
				\begin{equation}
				\begin{split}
				J_{twe}^c=\sum_{z^l \in Z=\lbrace z^h,z^d,z^w,z^m\rbrace}^{} \sum_{z_k^l \in z^l}^{}L_{\phi}(z_k^l) \\
				J_{twe}^d=\sum_{z^l \in Z=\lbrace z^h,z^d,z^w,z^m\rbrace}^{} \sum_{z_k^l \in z^l}^{}L_{\theta}(z_k^l)
				\end{split}				
				\label{loss-functions-final}				
				\end{equation}
		To conclude, we clarify the way we compute each of the twin attributes of the slab-based coefficiency. For the \textit{contiguity} model, we adopt the Jaccard coefficient \cite{hosseini2017leveraging} to assess how the word pairs collectively appear together in various slabs.
		As Eq. \ref{Jaccard-index-hour} shows, $w_i^{z_k^l}$ counts the number of instances where $w_i$ has been used in slab $k$ of temporal facet $l$, and $w_i^{z^l}$ represents the total number of usages for $w_i$ in dimension $l$, i.e. in any slab of dimension $l$.\\
		\vspace{-3mm}
		\begin{equation}
			\mathcal{F}_{con}(w_i,w_j,z_k^l)= \frac{\left | w_i^{z_k^l} \bigcap w_j^{z_k^l} \right |}{\left | w_i^{z^l} \bigcup w_j^{z^l} \right |}
			\label{Jaccard-index-hour}
			\vspace{-2mm}	
		\end{equation}
		For \textit{depth}($\mathcal{F}_{dep}(w_i,w_j,z_k^l)$), we are inspired by the concept of multi-aspect temporal influence \cite{hosseini2017leveraging}. While we include the effects from parent temporal latent facets, we declare (Eq. \ref{slab-based-coeff-function}) the depth property as the joint probability for each pair $(w_i,w_j)$ to mutually evolve in the same slab of $k$ in dimension $l$, i.e. $Pr(w_i,w_j,z_k^l)$. Based on subset rule $z^l \subset z^{l+1} \subset z^{l+2} \dots z^{t-1} \subset z^{t}$, where for instance, we consider the second slab in \textit{hour} latent facet, we will be required to further embrace the impacts from the slabs in the parent latent facets of \textit{day}, \textit{week} and \textit{month}.
		\vspace{-2mm}
		\begin{equation}
			\small
			\label{slab-based-coeff-function}
			\begin{split}
				Pr(w_i,w_j,z_k^l)\propto Pr(w_i) Pr_{\omega}(w_j|w_i)\quad\quad\quad\quad\quad\quad\quad  \\Pr(z_k^{l}|z^{\{l+1,l+2,\cdots,t\}},w_i,w_j)\quad\quad\quad
				\\Pr(z^{l+1}|z^{\{l+2,l+3,\cdots,t\}},w_i,w_j) \cdots\quad \\ Pr(z^{t-1}|z^{\{t\}},w_i,w_j) Pr(z^{t}|w_i,w_j)\quad
			\end{split}
			\vspace{-2mm}
		\end{equation}

		In general, since all the words are treated equally, we assign $Pr(w_i)$ by one. Also, $Pr_{\omega}(w_j|w_i)$ is the non-temporal weight for $w_j$ to co-occur in the context of $w_i$ that is computed using \textit{Collaborative Filtering}. Moreover, we assume that ${l+1}$ is the parent dimension of ${l}$, e.g. if $l$ is \textit{hour} latent factor, $l+1$ and $l+2$ will be respectively the \textit{day} and \textit{week} temporal facets. Hence, the unifacet temporal slab $z_k^l$ will be hierarchically impacted by each of the slabs in parent dimensions, i.e. $z^{l+1},z^{l+2},\dots,z^{t}$. Lastly, the log form of Eq. \ref{slab-based-coeff-function} substitutes the multiplication by summation in both sides of the equation. This can prevent the value in right-hand side to get demoted to less than digital minima. Note, that compilers, e.g. $C\#$ set values less than digital minima to 0.
		\vspace{-3mm}		
		\subsubsection{Multifacet Generative Model}
		\label{Multifacet-Generative-Model}
		\vspace{-1mm}		
		In this section, we devise a Multifacet Generative Model (MGM) which elaborately engages numerous temporal dimensions to infer the correlation weight ($\omega_{mgm}$) between each pair of the nodes in the propagation network. The model initially includes unlimited temporal facets ($z^1, z^2, \cdots  ,z^{t-1}$). Hence, the proposed model in this section can infer the \textit{time-only correlation} between the node pairs. Note that each temporal aspect, e.g. hour, is a latent factor and based on \textit{multifacet temporal dynamics} \cite{hosseini2018mining}, the joint probability for a pair of nodes ($n_i,n_j$) to spread a contagion is collectively inferred through how the propagation is occurred in multiple latent temporal facets (Eq. \ref{Step_000}). Accordingly, the weight is computed based on the behavior of the nodes in each of multi-facet temporal slabs ($\mathbb{\tau}^b$). Moreover, the proposed approach integrates the sequential time feature \cite{GomezRodriguez2010} \cite{Goyal2010} to further improve the accuracy of the edge weights.
		\vspace{-3mm}
		\begin{equation}
			\small
			\label{Step_000}
			Pr(n_i,n_j) \propto \sum\limits^{{\hbox{\,}}}_{\boldsymbol{{\sigma}}^{iterate}_{\{1,2,3,\dots,t\}} \in \mathbb{\tau}^b}  Pr(n_i,n_j,z)
			\vspace{-1mm}
		\end{equation}
		Intuitively, the temporal subset property \cite{hosseini2017leveraging} ($z^1 \subset z^2 \subset z^3 \cdots  z^{t-1} \subset z^t$) justifies that each temporal dimension is affected by its parent latent factor(s). For instance, as verbalized in Eq. \ref{Step_02}, given four latent factors of hour $z^h$, day $z^d$, week $z^w$, and month $z^m$, an outbreak in hour dimension may differ in various days, weeks, and months ($z^{\{d,w,m\}}$).
		\vspace{-2mm}
		\begin{equation}
		\small
		\label{Step_02}
		\begin{split}	
		Pr(z|n_i,n_j)=Pr(z^h|z^{\{d,w,m\}},n_i,n_j)\quad\quad\quad\quad\quad\quad\\
		Pr(z^d|z^{\{w,m\}},n_i,n_j) \quad\quad\quad\quad\quad\quad\\ Pr(z^w|z^{\{m\}},n_i,n_j)Pr(z^m|n_i,n_j)
		\end{split}
		\end{equation}
		
		Accordingly, Eq. \ref{Step_02_Gen} generalizes Eq. \ref{Step_02} to comprise infinite $t$ temporal dimensions.
					\vspace{-2mm}
					\begin{equation}
					\small
					\label{Step_02_Gen}
					\begin{split}	
					Pr(z|n_i,n_j)=Pr(z^1|z^{\{2,3,\cdots,t\}},n_i,n_j) \quad\quad\quad\quad\quad\quad\quad\quad\\Pr(z^{2}|z^{\{3,4,\cdots,t\}},n_i,n_j) \cdots\quad\quad\quad\quad\quad\;\;\\ Pr(z^{t-1}|z^{\{t\}},n_i,n_j)Pr(z^{\{t\}}|n_i,n_j)
					\end{split}
					\vspace{-2mm}
					\end{equation}										
					Basically, the multilateral joint probability for each node pair can be formalized by the probabilistic \textit{chain rule} Eq. \ref{Step_01_main}.					
					\vspace{-2mm}
					\begin{equation}
					\small
					\vspace{-2mm}
					\label{Step_01_main}	
					Pr(n_i,n_j,z) \propto Pr(n_i) Pr_{Seq}(n_j|n_i) Pr(z|n_i,n_j)
					\end{equation}
					The impact of all the nodes in propagation is the same (i.e. $Pr(n_i)=1$). Furthermore, we integrate the \textit{sequential temporal property} \cite{GomezRodriguez2010,Peng2017} to infer diffusions in each node $n_j$ given another node $n_i$ ($Pr_{Seq}(n_j|n_i)$). Hence, the proposed model in this section considers miscellaneous features including multi-aspect and sequential temporal properties. On the one hand, the computed edge weight counts how the nodes transfer the contagions during each of multi-facet temporal slabs ($Pr(z|n_i,n_j)$), and on the other hand, given the propagation threshold, the model infers how each node may spread a contagion to another ($Pr_{Seq}(n_j|n_i)$).					
					To conclude, we can determine the likelihood for each node pair to bilaterally get infected by the same contagion via replacing Eq. \ref{Step_02_Gen} into Eq. \ref{Step_01_main}, which results in the final Eq. \ref{Step_03_Gen}.					
					\vspace{-2mm}
					\begin{equation}
					\small
					\label{Step_03_Gen}
					\begin{split}
						Pr(n_i,n_j,z)\propto Pr(n_i) Pr_{Seq}(n_j|n_i) Pr(z^{1}|z^{\{2,3,\cdots,t\}},n_i,n_j)\\ \quad
						Pr(z^{2}|z^{\{3,4,\cdots,t\}},n_i,n_j) \cdots Pr(z^{t-1}|z^{\{t\}},n_i,n_j) Pr(z^{t}|n_i,n_j)
						\end{split}
							\vspace{-3mm}				
					\end{equation}

Finally, by substituting Eq. \ref{Step_03_Gen} into Eq. \ref{Step_000}, our recommended model can generally determine the degree of correlation between each pair of nodes ($n_i,n_j$). The log-likelihood from both sides of the Eq. \ref{Step_03_Gen} can avoid the outcome multiplication of the possible tiny-valued parameters to fall less than language defined digital minima, which is intrinsically the main cause for illegitimate zero-valued edge weights.
\vspace{-5mm}
\begin{equation}
\small
\label{Step_7}
\Xi (M)=\sum_{<n_i,n_j> \in <\mathbb{N},\mathbb{N}>}^{} Log(Pr(n_i,n_j;M))
\vspace{-2mm}
\end{equation}
Furthermore, referring to the industry use-case, the majority of the nodes raise scattered alarms in diverse times, which results in data insufficiency. To address this issue, our MGM model exploits temporal slabs to recompense the inner nodes missing diffusion paths. Moreover, given Eq. \ref{Step_03_Gen}, let $M$ denote the set of parameters and include $Pr(z^{1}|z^{\{2,3,\cdots,t\}},n_i,n_j)
, Pr(z^{2}|z^{\{3,4,\cdots,t\}},n_i,n_j), \cdots,\\ Pr(z^{t-1}|z^{\{t\}},n_i,n_j)$, and $Pr(z^{t}|n_i,n_j)$. Correspondingly, we can adopt the generalized Expectation-Maximization(EM) model which is proposed in \cite{hosseini2017leveraging} to maximize the log-likelihood of $\Xi(M)$ (Eq. \ref{Step_7}). In this way, we can also infer the best values for each parameter $\mu$ in $M$.
\vspace{-2mm}
		\vspace{-3mm}
		\subsection{Online Processing}		
		\vspace{-1mm}
		In the online phase, we firstly generate the Temporal-Textual Multi-attribute Graph which comprises diverse edges between each pair of the vertices in dynamical processes. Subsequently, the \textit{coherence score} determines how the trilateral edges of the multi-attribute graph can aggregate to form the latter single-edged \textit{Bilateral Network}. Eventually, we introduce a \textit{Max-Heap Graph cutting} algorithm to retrieve the final subgraphs with highly correlated nodes.		
		\vspace{-3mm}
		\subsubsection{Coherence Score and Bilateral Network}
		\vspace{-1mm}
				\label{Scoring-Propagation-Coherence}
				The \textit{Coherence Score} is decided via training in offline phase (Eq. \ref{MixtureModel-hybrid}) and collectively measures the final correlation weight between each pair of nodes. In this work, we consider three types of coherence: \textit{global semantical coherence} (Section \ref{original-word-embedding}), \textit{multi-aspect temporal-semantical balance} (Section \ref{time-aware-word-embedding}), and \textit{time-only correlation} (Section \ref{Multifacet-Generative-Model}).
				\vspace{-1mm}
				\begin{equation}
				\omega_{\textbf{C}}(u,v)=(1-\lambda-\beta)\times \omega_{twe}+\lambda\times\omega_{owe}+\beta\times\omega_{mgm}
								\vspace{-1mm}
				\label{MixtureModel-hybrid}
				\end{equation}
				As formulated in Eq. \ref{MixtureModel-hybrid}, we believe that two vertices in a propagation graph are tightly cohesive if three conditions hold. Firstly, the vector representation of their textual contents are similar, that is measured by Original Word Embedding ($\omega_{owe}$), Secondly, given associated textual contents of the contagions for the pair, the multi-aspect temporal co-occurrence pattern of the words, computed using Time-aware Word Embedding ($\omega_{twe}$), are highly correlated. Thirdly, the two vertices get infected to the same-typed contagions in homogeneous multi-aspect temporal facets, that is assembled through Multifacet Generative Model ($\omega_{mgm}$). Accordingly, we can either opt for a tuning procedure to learn the values for $\lambda$ and $\beta$ or borrow a \textit{page-rank} like model to automatically achieve the final edge weights that can also maximize the effectiveness in subgraph mining.
				Based on the acquired coherence score, we can consequently combine the three weights in a hybrid manner to retrieve the ultimate undirected Bilateral network. Note that there is only one edge between each pair of vertices in the bilateral network, i.e. no duplicate edges and no loops.
\vspace{-3mm}	
		\subsubsection{Max-Heap Graph-Cutting}
		\label{Max-Heap Graph-Cutting}
		Intuitively, exploiting subgraphs with highly correlated nodes requires \textit{canonical labeling} which is followed by numerous subgraph \textit{isomorphic tests} to assert that no subgraph can be returned twice or more. Nevertheless, the \textit{canonical labeling} algorithm is computationally declared as NP-hard problem. Furthermore, given multi-attribute graph, mining of the subgraphs with the highest edge weights necessitates to perform enumeration on every possible batch of nodes and decide whether any derived subgraph includes tied in edge weights or not, which further justifies why the problem is NP-hard. Consequently, since subgraph mining is typically regarded as an online continuous procedure and also an underlying phase in various tasks like preventative maintenance or immunization, we need to reduce the complexity of the proposed solution. Accordingly, our proposed framework initially transfers the subgraph mining problem to the computation of edge-weights between the pair of nodes.  Subsequently, the proposed \textit{Max-Heap graph-cutting} algorithm (Algorithm \ref{alg5}) can efficiently process the bilateral network $G$ to leverage the final set of tied in subgraphs.
		\vspace{-2mm}
				\begin{algorithm}[H]
					\caption{Max-Heap Graph Cutting (MaxHGC)}
					\label{alg5}
					\textbf{Input:} $G=(\mathbb{N},\mathbb{E}); W(\mathbb{E})=\{w(e)|e \in \mathbb{E}$\}\\
					\textbf{Output:} $G'=\left(\mathbb{N}',\mathbb{E}'\right)$
					\begin{algorithmic}[1]
						\STATE $\mathbb{N}'=\emptyset, \mathbb{E}'=\emptyset, \mathbb{E}"= \mathbb{E}, \mathbb{N}"= \mathbb{N}$
						\STATE $ MaxHeap\; H = \emptyset $
						\WHILE{$\mathbb{E}" \neq \emptyset$}
						\STATE add $e=(u,v) \in \mathbb{E}" \;into\; H\;based\;on\;w(e)$
						\STATE remove $e$ from $\mathbb{E}"$
						\ENDWHILE
						\WHILE{$\mathbb{N}" \neq \emptyset$}
						\STATE remove $e=(u,v)$ with\;maximum $w(e)$ from $H$
						\STATE $Sort(H)$
						\IF{$u | v \in \mathbb{N}"$}
						\STATE add $e=(u,v)$ into $\mathbb{E}'$
						\STATE remove $u,v$ from $\mathbb{N}"$
						\ENDIF
						\IF{$\;u,v\;\notin\;\mathbb{N}'$}
						\STATE add $u,v$ into $\mathbb{N}'$
						\ENDIF
						\ENDWHILE
						\STATE return $G'$
					\end{algorithmic}					
				\end{algorithm}
				\vspace{-4mm}
		Algorithm \ref{alg5} runs as follows: First, given $G=(\mathbb{N},\mathbb{E})$ as the bilateral network, the probability for every pair $e = (u, v)$ will be proportional to the edge weight between them. Correspondingly, the bigger the weight, the higher the probability for the pair to appear in the same subgraph. As the next step, we duplicate $\mathbb{N}$ and $\mathbb{E}$, as the respective set of nodes and edges, into $\mathbb{N}''$ and $\mathbb{E}''$ to maintain the primary bilateral network $G$ unaffected. After making the primary Max Heap $H$ empty, given the weight function $w(e)$, it adds each edge $e \in \mathbb{E}''$ into $H$ and removes the edge from $\mathbb{E}''$ afterward. Subsequently, it removes the edge $e$ with the largest weight from the heap $H$, Reheapifies $H$, adds the corresponding nodes of $e$ to $\mathbb{N}'$, and removes the nodes from $\mathbb{N}''$. The eliminated edge $e$ from the max-heap is then added to $\mathbb{E}'$. This process is repeated until no nodes exist in $\mathbb{N}''$. Finally, it returns the graph $G'$ with an optimized number of nodes $\mathbb{N}'$ and edges $\mathbb{E}'$, where each group of connected nodes can form an exploited subgraph. In a nutshell, Firstly, the procedure in algorithm \ref{alg5} ensures that all the edges in $G'$ are among the top $k$ highest weights from the input bilateral graph $G$. Secondly, the resulting graph $G'$ contains all the nodes of $G$, and eventually maintains a minimum number of required edges. Therefore, the proposed max-heap graph-cutting algorithm optimizes the number of exploited subgraphs and maximizes the intra-subgraph correlations. In other words, it maximizes the correlation inside each of exploited subgraphs, and at the same time, minimizes the number of edges that are utilized in subgraph mining.
		\vspace{-5mm}
		\section{Experiments}
		\label{Experiments}
		\vspace{-1mm}
		We conducted extensive experiments on two real-world datasets to examine the performance of our approach in mining of the subgraphs with tightly correlated nodes. We developed our algorithms using C\#, Python, and Microsoft Structured Query Language. Aiming to reduce the number of database queries, we utilized the .NET Language-Integrated Queries (LINQ). The experiments were executed on a server with 4.20GHz Intel Core i7-7700K CPU and 64GB of RAM. We publicly share\footnote{https://sites.google.com/view/time-aware-embedding} the \underline{code} and \underline{data}.
		\vspace{-5mm}
		\subsection{Data}
		\label{Data}
		\vspace{-2mm}
		 We used two real-life datasets (ST-0 and ST-1) for experiments. The datasets, collected in 2016, report how the faults are propagated between the nodes in each of two Singaporean train stations. The contagions in each of datasets are propagated over more than 1 million nodes, where each node is at most associated with up to 300 contagions.
		\vspace{-4mm}
		\subsection{Benchmark}
		\label{benchmark}
		\vspace{-1mm}
Intuitively, we define statistical hypothesis parameters to evaluate the effectiveness of our proposed framework in subgraph mining. In summary, we investigate whether all the nodes in each discovered subgraph are correctly chosen, and no missing node is left out. 
To setup, we dedicate 80\% of each dataset for parameter setting and accomplish the test on the remaining 20\%.
Take $n_i$ as a node in a detected subgraph $g_i$ and presume that the contagion is transmitted if the type of different contagions is the same as $y_i$. For instance, influenza can be transferred by three categories of A, B, and C, but actually, all of them are of the same disease. We can define the statistical parameters as follows:\\
\textit{True Positive} is observed when $n_i$ is reported to get infected by a contagion $c_i$ of type $y_i$ in time $t$ and at least one of its detected neighbors in the subgraph receives a contagion of the same type $y_i$ during period of ($t-\theta_t,t+\theta_t$) where the model also estimates that $n_i$ has got infected by $y_i$ typed contagion. \textit{False Positive} can, likewise, represent the number of cases when $n_i$ does
			not raise the alarm of type $y_i$ and at least one
			of its neighbors signifies the alarm of the type $y_i$
			during the temporal period, but in fact the model
			asserts that $n_i$ raises the alarm of the type $y_i$.
			Using similar logics we can also define \textit{True Negative} and \textit{False Negative}.
		\noindent Given four parameters of TP, FP, TN, and FN, we can calculate performance metrics of Precision, Recall, and F-measure. We 
		determine the best model by F-measures.
		\vspace{-5mm}
		\subsection{Effectiveness}
		\vspace{-1mm}
		In this section, we elucidate the impact of influencing parameters on our proposed framework. We then report how our approach overcomes the performance of other rivals.
		\vspace{-3mm}
		\subsubsection{Effectiveness of Big and Small Clusters}
		\label{Impact-of-zeta-on-effectiveness-of-Big-and-Small-Clusters}
		\vspace{-1mm}
Given the textual contents of the contagions, the Time-aware Word Embedding ($\omega_{twe}$) module of our framework takes the temporal word-word co-occurrence patterns into consideration to infer the correlation weight between each pair of nodes. One prerequisite of time-aware word embedding is to utilize the similarity grids (Sec. \ref{Constructing-Similarity-Grids}) to extract the multi-aspect temporal slabs (Sec. \ref{Acquiring-Temporal-Slabs}). In a nutshell, clustering of the highly similar splits in $t$ temporal dimensions constructs the slabs and reduces the sparsity in $\mathbb{NTA}$ cube. Nevertheless, the threshold of the agglomerative clustering is quite important as it can integrate mutually exclusive splits into the same slab or assign related splits to separate clusters. As explained in Section \ref{Acquiring-Temporal-Slabs}, the clustering generates two series of slabs, namely, \textit{small} and \textit{big clusters}. Hence, for each dataset, we train the most suitable time-aware embedding approach (depth or contiguity regression models as verbalized in Section \ref{time-aware-word-embedding}), and compare the performance metric of big and small clusters as defined in Section \ref{benchmark}.
\vspace{-3mm}
\begin{table}[H]
	\centering	
	\caption{Effectiveness of Big and Small Clusters}	
	\vspace{-3mm}
	\label{Effectiveness-of-Big-and-Small-Clusters}
	\begin{tabular}{c|c|c|c|c|}
		\cline{2-5}
		& \multicolumn{2}{c|}{ST-0}                                                                                         & \multicolumn{2}{c|}{ST-1}                                                                                         \\ \hline
		\multicolumn{1}{|c|}{$\zeta$}    & \begin{tabular}[c]{@{}c@{}}Small\\ Clusters\end{tabular} & \begin{tabular}[c]{@{}c@{}}Big\\ Clusters\end{tabular} & \begin{tabular}[c]{@{}c@{}}Small\\ Clusters\end{tabular} & \begin{tabular}[c]{@{}c@{}}Big\\ Clusters\end{tabular} \\ \hline
		\multicolumn{1}{|c|}{5}          & 0.5850 & \textbf{0.5956}                                     & 0.5333                                                   & \textbf{0.5420}                                        \\ \hline
		\multicolumn{1}{|c|}{10}         & \textbf{0.5669}                                       & 0.5589                                              & \textbf{0.5564}                                          & 0.5561                                                 \\ \hline
		\multicolumn{1}{|c|}{15}         & \textbf{0.5681}                                       & 0.5431                                              & 0.5454                                                   & \textbf{0.5635}                                        \\ \hline
		\multicolumn{1}{|c|}{20}         & 0.5381                                                & \textbf{0.5501}                                     & 0.5458                                                   & \textbf{0.5732}                                        \\ \hline
		\multicolumn{1}{|c|}{25}         & 0.5302 & \textbf{0.5489}                                     & 0.5575                                                   & \textbf{0.5831}                                        \\ \hline
		\multicolumn{1}{|c|}{30}         & \textbf{0.5404}                                         & 0.5342                                              & 0.5490                                                   & \textbf{0.5959}                                        \\ \hline
		\multicolumn{1}{|c|}{35}         & 0.5448 & \textbf{0.5589}                                     & 0.5676                                                   & \textbf{0.6048}                                        \\ \hline
		\multicolumn{1}{|c|}{40}         & 0.5083                                                & \textbf{0.5228}                                     & 0.5643                                                   & \textbf{0.6077}                                        \\ \hline
		\multicolumn{1}{|c|}{45}         & 0.5211                                                & \textbf{0.5682}                                     & 0.5719                                                   & \textbf{0.6218}                                        \\ \hline
		\multicolumn{1}{|c|}{50}         & 0.5146                                                 & \textbf{0.5603}                                     & 0.5755                                                   & \textbf{0.6174}                                        \\ \hline
		\multicolumn{1}{|c|}{Comparison} & 3                                                        & 7                                                      & 1                                                        & 9                                                      \\ \hline
	\end{tabular}
	\vspace{-4mm}
\end{table}

Table \ref{Effectiveness-of-Big-and-Small-Clusters} compares the effectiveness of $\omega_{twe}$ using small and big clusters. Moreover, the experiment illustrates the impact of including different top $\zeta$ items from the vector representation of the words in textual contents of contagions. From Table \ref{Effectiveness-of-Big-and-Small-Clusters} we can observe that for most $\zeta$ values, the big clusters can better obtain the highly connected subgraphs than the small clusters. Consequently, we opt for big clusters to maximize the effectiveness of the $\omega_{twe}$ module. From another perspective, we observe that the performance metric of the time-aware approach in ST-1 is better than ST-0. The reason is that the ST-1 dataset is denser than ST-0 and includes a bigger number of spreads in various time-stamps. On the other hand, the propagations in ST-1 comprehensively occurred in various temporal slabs of different dimensions.
\vspace{-5mm}
		\begin{figure}[H]
			\centering
			\minipage{0.49\linewidth}
				\centering
				\includegraphics[width=1\textwidth]{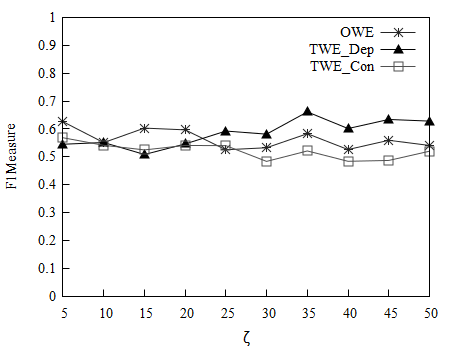}
				\vspace{-2mm}
				\small ST-0 F-measure Changes
				\label{fig:F1Measure8MeasureBig}
			\endminipage
			\centering
			\minipage{0.49\linewidth}
				\centering
				\includegraphics[width=1\textwidth]{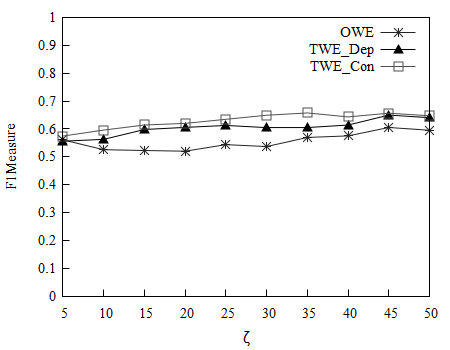}		
				\vspace{-2mm}
				\small ST-1 F-measure Changes
				\label{fig:F1Measure10MeasureBig}
			\endminipage
			\hfill
			\vspace{-1mm}
			\caption{Effectiveness of embedding models}
			\label{fig:F1MeasurePlotsBig}
		\end{figure}
		\vspace{-7mm}
		\subsubsection{Impact of $\zeta$ on Embedding}
		\label{Impact-of-zeta-Embedding}
		\vspace{-1mm}				
		Intuitively, either of the slab-based coefficient attributes (\textit{depth} or \textit{contiguity}) can possibly be observed in a dataset. From one perspective, the word-word co-occurrence patterns in a temporal slab may get influenced by the \textit{parent} temporal slab(s) (depth attribute). For instance, the number of trips to the downtown in an hour can differ in various days. Here the hour dimension is affected by its parent dimension (i.e. day). From another perspective, the occurrence pattern of the word pairs in the slab may neglect the impact from parent temporal aspect (contiguity attribute). The amount of shopping for the essential requirements of the families is approximately the same during the weekdays and weekend. Hence, to address the perturbations, as elucidated in Section \ref{time-aware-word-embedding}, we propose two distinctive weighted least square regression models to delineate the contiguity ($TWE\_{Con}$) and depth ($TWE\_{Dep}$) attributes. Furthermore, we compare the time-aware regression models with the state-of-the-art original word embedding ($OWE$).\\
		We also incorporate the effect of the top $\zeta$ items from word vectors in the performance analysis of subgraph mining.		
		Note that with regard to the parameter setting for OWE, we use the same approach in \cite{Pennington2014}. Thus, the value of $\alpha$ is assigned by $\frac{3}{4}$ and $x_{max}$ equates to 100. As Fig. \ref{fig:F1Measure8MeasureBig} depicts the best performance in ST-0 dataset is achieved by $TWE\_{Dep}$ at $\zeta=35$. Similarly, for ST-1 dataset, the best effectiveness in exploiting of the highly correlated subgraphs is achieved by the contiguity-based time-aware approach $TWE\_Con$ at $\zeta=35$.
		On the one hand, $TWE\_Con$ overcomes other models in ST-1 dominantly (for all $\zeta$ values), and on the other hand, for ST-0, the $TWE\_Dep$ proves superiority over both OWE and $TWE\_Con$ where the word vectors are represented by greater than 20 items ($>20$). It is worth noting that the range [$25,50$] in both datasets can reasonably demonstrate good performance in vector representation.
		\vspace{-3mm}		
		\subsubsection{Impact of Embedding initialization on accuracy}
		\vspace{-1mm}
		Intuitively, the embedding cost function aims to minimize the error between the co-occurrence of the word pairs (e.g. ($w_i$,$w_j$)) and the inner product of their corresponding main and context vectors, denoted by $(\vec{w}_i^m)^T$ and $(\vec{w}_j^c)$. However, the word embedding approaches \textit{randomly initialize} the primary vector space for each instance word $w_i$. As explained in Section \ref{original-word-embedding}, the primary random vector is important as it can regulate the final biases, i.e. $b_i^m$ and $b_j^c$.\\
		Surprisingly, the randomly initialized matrix (whose vectors are associated with each of the words) results in a different local maxima every time that the embedding procedure is executed. Therefore, unlike \textit{inheritance algorithms} which perform \textit{mutation} to avoid local maxima, the word-embedding models turn out to be partially dependent on the primary matrix which can negatively affect the performance of the framework and result in a \textit{marginal error}. What is described to put the validity of the \textit{count-based} Neural word embedding models (e.g. GloVe) under strict scrutiny. Since we, likewise, integrate a count-based time-aware embedding module into our framework, we logically need to examine the unsigned size of the marginal error using the easy to interpret \textit{Mean Absolute Percent Error} (\textit{MAPE}), Eq. \ref{MAPE}.
		\vspace{-2mm}
						\begin{equation}
						\label{MAPE}
						MAPE= [ \frac{1}{n} \sum_{}^{} \frac{|Actual - Forecast|}{Actual} ] \times 100				
						\vspace{-3mm}
						\end{equation} 
		In practice, where we subsequently run both original and time-aware models ($TWE\_Con$ and $TWE\_Dep$) for 10 rounds (denoted by $n$ in Eq. \ref{MAPE}), we estimate the \textit{Actual} effectiveness by the \textit{mean} and \textit{median} of all F-measures.
		Correspondingly, we assign the \textit{Forecast} by the F-measure in each round of execution. Finally, the MAPE is signified by the average of the deviation size between actual and Forecast metrics. Table \ref{Accuracy-change-initial-random-matrix} presents the results of the experiment on two real datasets. To gain the fairness, for each competitor model, the setting for $\zeta$ that can maximize the subgraph mining effectiveness has been applied.\\
		As evident in Table 6, the average rate of error (MAPE) caused by the random initialization is remarkably inflated.				
					Regardless of the fact that either MAPE rate for our model is less ($TWE\_Con$ in ST-1) or more ($TWE\_ Dep$ in ST-0) than the basic baseline (OWE), the outcome effectiveness of our proposed time-aware approaches can still overcome OWE.	
																			\vspace{-3mm}
																			\begin{table}[htp!]
																				\centering
																				\caption{Accuracy Range After Iterative Runs}
																				\vspace{-3mm}
																				\label{Accuracy-change-initial-random-matrix}
																				\begin{tabular}{|c|c|c|c|c|}
																					\hline
																					Station               & \begin{tabular}[c]{@{}c@{}}Method \end{tabular} & $\zeta$             & \begin{tabular}[c]{@{}c@{}}Forecast\\ value\end{tabular} & \begin{tabular}[c]{@{}c@{}}Accuracy range \\ with MAPE\end{tabular} \\ \hline
																					& \multirow{2}{*}{$OWE$}                                     & \multirow{2}{*}{5}  & Mean                                                     & 0.57814 $\pm$ 0.03467                                       \\ \cline{4-5} 
																					\multirow{2}{*}{ST-0} &                                                         &                     & Median                                                   & 0.57529 $\pm$ 0.03415                                           \\ \cline{2-5} 
																					& \multirow{2}{*}{$TWE\_Dep$}                                     & \multirow{2}{*}{35} & Mean                                                     & 0.65797 $\pm$ 0.05178                                       \\ \cline{4-5} 
																					&                                                         &                     & Median                                                   & 0.64383 $\pm$ 0.05092                                          \\ \hline
																					& \multirow{2}{*}{$OWE$}                                     & \multirow{2}{*}{45} & Mean                                                     & 0.57248 $\pm$ 0.03932                                      \\ \cline{4-5} 
																					\multirow{2}{*}{ST-1} &                                                         &                     & Median                                                   & 0.58261 $\pm$ 0.03706                                           \\ \cline{2-5} 
																					& \multirow{2}{*}{$TWE\_Con$}                                     & \multirow{2}{*}{35} & Mean                                                     & 0.58701 $\pm$ 0.02369                                      \\ \cline{4-5} 
																					&                                                         &                     & Median                                                   & 0.58611 $\pm$ 0.02350                                         \\ \hline
																				\end{tabular}
																				\vspace{-6mm}
																			\end{table}	

		\subsubsection{Effectiveness of subgraph mining}
		\label{effSubgraphMining}
\begin{figure}[H]
	\centering
	\minipage{0.40\linewidth}
		\centering
		\includegraphics[width=1\textwidth]{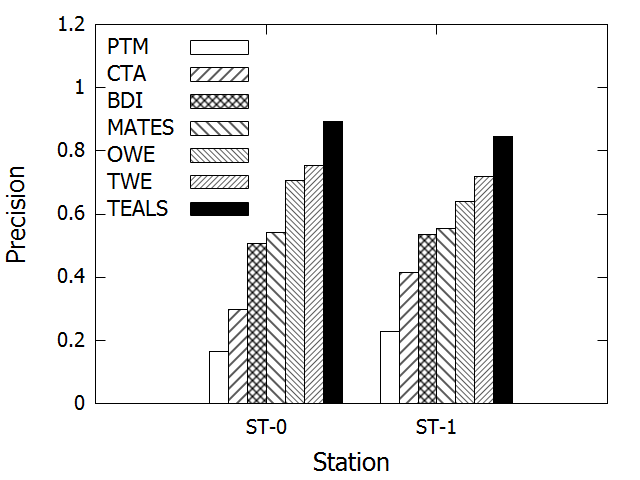}
		\vspace{0mm}
		\small Precision
		\label{fig:PrecisionComparison}
	\endminipage
	\centering
	\minipage{0.40\linewidth}
		\centering
		\includegraphics[width=1\textwidth]{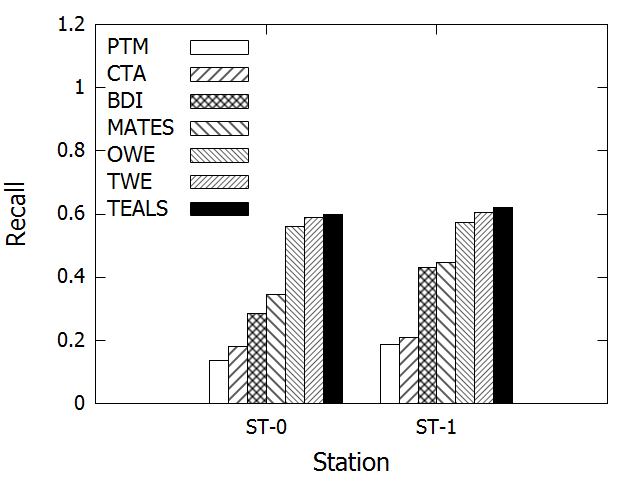}
		\vspace{0mm}
		\small Recall
		\label{fig:RecallComparison}
	\endminipage \\
	\minipage{0.40\linewidth}
		\centering
		\includegraphics[width=1\textwidth]{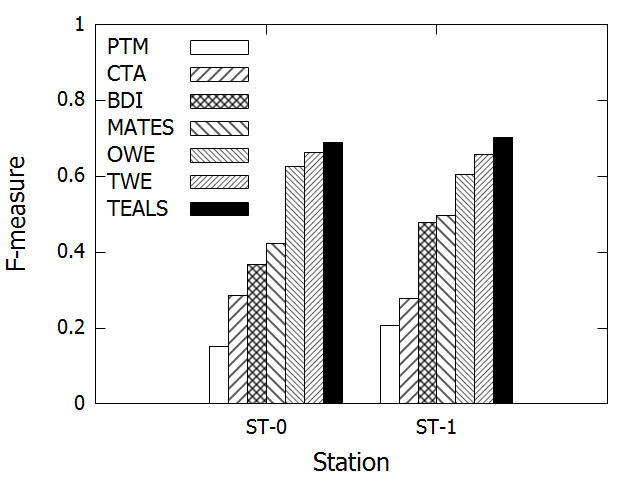}
		\vspace{0mm}
		\small F-measure
		\label{fig:RecallComparison}
	\endminipage
	\vspace{-3mm}
	\hfill
	\caption{Effectiveness of subgraph mining}
	\vspace{-4mm}
	\label{fig:MethodsComparison}
\end{figure}

		In general, the problem regarding mining of the subgraphs with highly correlated nodes can be mitigated to the so-called edge weight calculation problem.
		In other words, the subgraph mining models typically run in two steps: First, they compute the edge weights between the nodes; Second, they extract the subgraphs using a graph-cutting algorithm. We now employ the benchmark (Section \ref{benchmark}) to examine the performance of seven algorithms in exploiting of the tightly correlated subgraphs:
		\vspace{-2mm}		
		\begin{itemize}
		\item \textit{Pair-wise Temporal Model} (PTM, WSDM-2010): the model employs sequential temporal property \cite{Peng2017,Prakash2010,GomezRodriguez2010,Goyal2010} to infer edge weights. 
		\item \textit{Concurrent Temporal Approach} (CTA, WWW-2017): the generative model \cite{hosseini2017leveraging} infers propagation in two dimensions of the hour and day.
		\item \textit{Bidirectional Diffusion Inference} (BDI): the bidirectional version of diffusion inference model \cite{GomezRodriguez2010} (KDD-2010) which  detects the cascades in quadrilateral dimensions \cite{hosseini2017leveraging} of the hour, day, week and month.						
		\item \textit{Multi-Attribute Time trends to Exploit Subgraphs} (MATES): the unsupervised hybrid model adjusts the sequential and Multi-Aspect temporal parameters to learn the edge weights.
		\item \textit{Original Word Embedding} (OWE, EMNLP-2014): attains semantical links between nodes via GloVe \cite{Pennington2014}.
		\item \textit{Time-aware Word Embedding} (TWE): our propositional algorithm that is elucidated in Section \ref{time-aware-word-embedding}.
		\item \textit{\textbf{T}emporal-\textbf{T}extual \textbf{E}mbedding \textbf{A}pproach to \textbf{L}everage \textbf{S}ubgraphs} (TEALS): our framework in Section \ref{Framework-Overview}.
	\end{itemize}								
	\vspace{-2mm}
Figure \ref{fig:MethodsComparison} compares the effectiveness of various competitors in subgraph mining. For temporal models, we believe BDI overcomes CTA because firstly it incorporates more latent temporal facets (four dimensions) than CTA (two latent factors of the hour and day) and secondly it specifically uses the cascades to glean the diffusions in both directions. Moreover, PTM gains the least performance as it only employs the sequential temporal property. Recall that the sequential spread of contagions can change in various temporal layers, e.g. the propagation between a pair of nodes may differ during the weekdays and weekends. Nevertheless, on the contrary, since MATES fuses both sequential and quadrilateral temporal properties, compared to other time-only approaches (PTM, CTA, and BDI), it achieves a better F-measure in subgraph mining.
Interestingly, the textual approaches gain a better performance compared to the temporal models. Even the most straightforward original word embedding approach overcomes the best temporal model (MATES). This explains that the correlation between the textual semantics of the nodes deems to be more authoritative than the singular temporal data. However, since TWE improves the OWE with the temporal techniques (i.e. quadrilateral temporal impact), even though the amount of improvement for recall metric is less than Precision, TWE shows its evident superiority over OWE in F-measure.\\
\vspace{-7mm}
		Note that given each pair of vertices ($u$,$v$), our framework (TEALS) (Section \ref{Scoring-Propagation-Coherence}) collaboratively integrates three weights of \textit{global semantical coherence} ($\omega_{owe}$), \textit{multi-aspect temporal-semantical balance} ($\omega_{twe}$), and \textit{time-only correlation} ($\omega_{mgm}$) to learn the final edge weight. Hence, aiming to demonstrate the impacts, as reported in Table \ref{framworkMetrics}, we train the best values for $\lambda$ and $\beta$ to maximize the effectiveness of the proposed framework. Nevertheless, we fuse the triad weights via a \textit{page rank} like algorithm. Most notably, while the amount of improvement in the recall is less than precision, TEALS gains a substantial improvement in F-measure.		
				\vspace{-6mm}
				\begin{table}[H]
					\centering
					\caption{ Mixture Modal Optimized Values}
					\vspace{-3mm}
					\label{framworkMetrics}
					\begin{tabular}{|c|c|c|}
						\hline
						\multirow{2}{*}{} & \multicolumn{2}{c|}{F-measure} \\ \cline{2-3} 
						& $\lambda$       & $\beta$       \\ \hline
						ST-0              & 0.2            & 0.4           \\ \hline
						ST-1              & 0.1            & 0.6           \\ \hline
					\end{tabular}
				\end{table}
				\vspace{-3mm}				
		Furthermore, as noticed in both datasets, $\omega_{owe}$ module plays a less significant role in the framework (0.2 in ST-0 and 0.1 in ST-1). This leads to larger influences for time-aware modules ($\omega_{twe}$ and $\omega_{mgm}$). On the other hand, since the sum of coefficients for time-aware modules are respectively 0.8 and 0.9 in ST-0 and ST-1, we can admit that both datasets are time-intensive. From another perspective, the performance metrics of the embedding models (OWE and TWE) are considerably higher than time-only approaches. Hence, given the task of subgraph mining in dynamical processes, the context semantics appear to be more dominant than the initial temporal features.
		\vspace{-3mm}
		\subsubsection{Effectiveness faces data variation}
\begin{figure}[H]
					\centering
					\minipage{0.44\linewidth}
						\centering
						\includegraphics[width=1\textwidth]{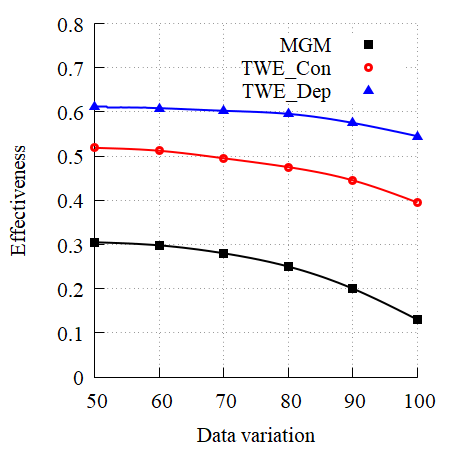}
						\vspace{0mm}
						\small Dataset ST-0
						\label{fig:impactSimilarityEffectivnessST-0}
					\endminipage
					\centering
					\minipage{0.44\linewidth}
						\centering
						\includegraphics[width=1\textwidth]{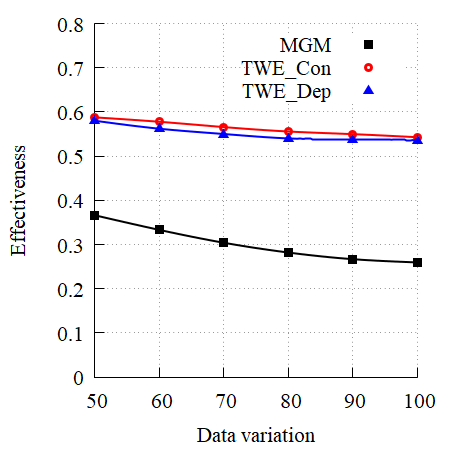}
						\vspace{0mm}
						\small Dataset ST-1
						\label{fig:impactSimilarityEffectivnessST-1}
					\endminipage
					\vspace{-3mm}
					\hfill
					\caption{F-measure of subgraph mining when dataset grows}
					\vspace{-4mm}
					\label{fig:Impact-constant-similarity-girds-data-variation-on-effectiveness}
				\end{figure}

		\vspace{-1mm}
		One of the most notable points about our framework is that the changes in similarity grids can initially affect the output temporal slabs and the credibility of subgraph mining process afterward. Since three time-aware models of MGM, TWE\_Con, and TWE\_Dep use the similarity grids, we need to estimate how frequently the similarity grids may evolve in time. As explained in Section \ref{Acquiring-Temporal-Slabs}, we divide the data in half and attain both temporal and textual similarity grids and measure the half-size (i.e. 50\%) effectiveness in exploiting of the subgraphs with highly correlated nodes. More specifically, where we repeatedly increase the size of the data by $\delta$ (set to 10\%) and apply the same similarity grids, we can experimentally study how the effectiveness changes in each interval. This will help us better adjust the Trigger interim which regenerate the similarity grids. Here we aim two goals: Firstly, to minimize the number trigger runs, and Secondly, to maximize the effectiveness of the subgraph mining process. Figure \ref{fig:Impact-constant-similarity-girds-data-variation-on-effectiveness} shows how the effectiveness of the time-aware modules are negatively affected where the data size grows and the similarity grids remain untouched.
		\vspace{-4mm}	
				Due to the bigger size of the corpus which leads to better training results, compared to ST-0, the effectiveness decreases less in dataset ST-1. As depicted, when the dataset expands, the models built on top of textual similarity grids (TWE\_Con and TWE\_Dep) demonstrate more robustness versus temporal grid approach (MGM). The reason is that while the temporal similarity grids change exquisitely by the sudden temporal skews, a limited number of words are only added to the corresponding co-occurrence matrices in TWE models. Furthermore, TWE models collectively gain the contextual support of the text and time, which help them better tolerate the change spikes. However, while TWE\_Con employs the Jaccard function to compute the surface correlation between each pair of nodes, TWE\_Dep, employs the Bayes theory through a comprehensive set of miscellaneous parameters, which can justify why the performance of TWE\_Dep shrinks less in the course of time.\\
				Nevertheless, Figure \ref{fig:Impact-constant-similarity-girds-data-variation-on-effectiveness} reflects how we can estimate the value for trigger interval ($\Delta$). For instance, if we set the threshold $\epsilon$ to 3\% in ST-0, we will need to reinitialize both temporal and textual similarity grids when the size of the dataset grows by 10\%. Similarly, for dataset ST-1, if we set the $\epsilon$ for temporal similarity grid to 4\%, we will be required to regenerate the corresponding similarity grids for MDM and TWE (both TWE\_Con and TWE\_Dep) models after 10\% and 30\% of growth ($\Delta$) respectively. As a result, the number of trigger runs for temporal similarity grids consumed by the MGM will be three times bigger than the number of calls for the TWE triggers. Finally, we can convert the percentage metric of $\Delta$ to its corresponding temporal period ($\Delta_t$). This will consequently impose that the trigger must be executed after each $\Delta_t$ interim.
		\vspace{-4mm}
		\subsection{Efficiency}
		\vspace{-1mm}
		Our framework exploits highly correlated subgraphs from propagation networks that is regarded as an online procedure that fosters applications like preventive medicine and preventative maintenance. Such applications require to continuously process millions of propagation events, which makes the time requirement of our framework eminently critical. Hence, in this section, we study the efficiency of our proposed algorithms.
		\vspace{-3mm}
		\subsubsection{Complexity Comparison}
		\vspace{-1mm}
		This section compares the time complexity of the original word embedding (OWE) versus our time-aware approach (TWE) from a theoretical computer science perspective. Intuitively, OWE consumes the complete set of vocabularies as a single slab which is denoted by $\mathbb{V}$. Let $\gamma$ be a constant multiplier and $\iota$ as the number of convergence iterations. In retrospect, it is quite obvious that OWE runs in $O(\gamma \times \iota \times |\mathbb{V}|^2)$ where the total number of vocabularies is denoted by $|\mathbb{V}|$. However, TWE divides the data into several smaller set of vocabularies where each of them corresponds to a distinct unifacet temporal slab. Remember that we propose a model to acquire temporal slabs in Section \ref{Acquiring-Temporal-Slabs}. Let $\mathbb{V}_k^l$ represent the vocabularies in textual contents of the slab $k$ of the latent facet $l$. The expected time for TWE to run each slab will thus yield in $O(\gamma \times \iota \times |\mathbb{V}_k^l|^2)$. Correspondingly, as Formula \ref{TWE_complexity} shows, the time complexity for CPU-based TWE will be collectively calculated through all the slabs. For simplification, we limit the latent space to four dimensions (the value of $\gamma$) where $z^m, z^w, z^d,$ and $z^h$ represent the month, week, day, and hour latent facets.
		\vspace{-3mm}
		\begin{equation}
			\label{TWE_complexity}
			\gamma \times \iota \times \sum_{z^l \in Z=\lbrace z^m,z^w,z^d,z^h\rbrace}^{} \sum_{z_k^l \in z^l}^{}|\mathbb{V}_k^l|^2
			\vspace{-2mm}
		\end{equation}
		Accordingly, due to the subtle connection between the size of the co-occurrence matrix and the time complexity, compared to OWE, TWE comes with two strong assurances in \textit{efficiency}: First, each slab in TWE can be processed in parallel with other slabs that fosters parallelism. Second, compared to the massive size of co-occurrence matrix in OWE, TWE works on a much smaller set of slab-based matrices.\\
		\textbf{Scalability.} as an interesting note, all methods are fast when the dataset is small, e.g. $|\mathbb{V}|$ = 10k. Nevertheless, in big data scenario where the number of vocabularies increases, the size of the input co-occurrence matrix will swell. As a result,  the performance of OWE will negatively get affected. In contrast, when TWE faces the large-scale data, we observe that compared to the total number of vocabularies, the ratio of the shared vocabularies between temporal slabs mitigates, so-called \textit{slab-vocabulary independence phenomena}. Consequently, the relative size of the slab-based matrices will decrease, which directly reduces the number of entries in slab-based co-occurrence matrices.
		\vspace{-4mm}
		\subsubsection{Impact of $\zeta$ on efficiency}
		\vspace{-1mm}
		 The semantic representation of each node can continuously evolve by arising of the new contagions.
Also note that the time-aware word embedding approaches should regenerate the word vectors after each trigger run. Nevertheless, we also need to know how the efficiency of embedding approaches change when each word $w_i$ in textual contents of sample node $u$ ($O_u$) is replaced with the top $\zeta$ most similar words from encyclopedic semantic representation of the node $O^\prime_u$. To this end, we combine both datasets (Section \ref{Data}) to study the impact of $\zeta$ on efficiency of word embedding algorithms: $OWE$, $TWE\_Dep$, and $TWE\_Con$.
		\vspace{-4mm}
		\begin{figure}[H]
			\centering
			\includegraphics[width=0.25\textwidth]{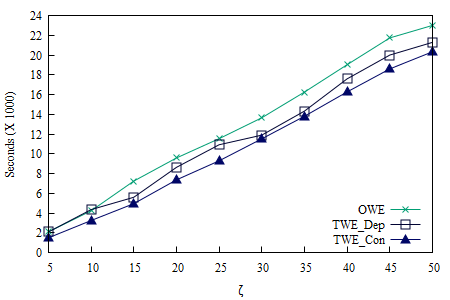}
			\vspace{-3mm}
			\caption{Impact of $\zeta$ on efficiency of semantic representation}
			\vspace{-4mm}
			\label{fig:EntitySimilarity1}
		\end{figure}
		Figure \ref{fig:EntitySimilarity1} depicts that as the value for $\zeta$ increases, the latency of all three methods slightly grows. The reason is that choosing a bigger value for $\zeta$ increases the size of the semantic representation of each vertex (e.g. $O^\prime_u$) which results in higher processing times. In general, $OWE$ as the original count-based model enumerates how frequently each instance word appears in a context and constructs the co-occurrence matrix afterward. Nevertheless, since the size of the non-temporal co-occurrence matrix in OWE is significantly larger than the slab-based temporal matrices in TWE\_Con and TWE\_Dep, operations on OWE matrix considerably take bigger times than time-aware approaches. From another perspective, as TWE\_Dep utilizes an extra step to infer the optimized values for the Bayesian parameters, it requires more processing time compared to $TWE\_Con$. Nevertheless, the less processing time does not necessarily lead to higher effectiveness, as in terms of effectiveness, TWE\_Dep outperforms the TWE\_Con in ST-0.\\
				\textbf{Parallelization}. 
				It is noteworthy that compared to OWE, our time-aware word embedding approaches ($TWE\_Con$ and $TWE\_Dep$) can be parallelized more conveniently. Accordingly, each core can process the comprising data in each distinguished temporal slab, where, in a GPU manner, the amount of final cost for the loss function will be collectively computed in all temporal facets ($z^1,z^2,\dots,z^t$).
\vspace{-5mm}
		\section{Conclusion}
		\vspace{-2mm}
		\label{Conclusion}		
		In this work, we devise a unified framework to effectively exploit a set of subgraphs with highly correlated nodes from propagation networks. More specifically, we divide
		the task of subgraph mining into two subtasks: edge weight computation and graph-cutting procedure. In summary, we firstly endorse a multi-aspect temporal generative model to compute the edge weights between the nodes in correlation networks.
		Secondly, we develop a neural network based time-aware
		word embedding algorithm which infers the co-occurrence patterns under an infinite number of temporal dimensions.
		Finally, we employ our max-heap graph-cutting algorithm to optimize the number of exploited subgraphs.
		The experimental results show that our proposed	framework outperforms existing state-of-the-art approaches in the
		field of subgraph mining from diffusion networks. 
		\vspace{-4mm}
		\ifCLASSOPTIONcaptionsoff
		\newpage
		\fi
		
		
		
		\vspace{-1mm}
		\bibliographystyle{IEEEtran}
		\bibliography{IEEEexample}
		\vspace{-14mm}
\begin{IEEEbiography}[{\includegraphics[width=1.0in,height=1.25in,clip,keepaspectratio]{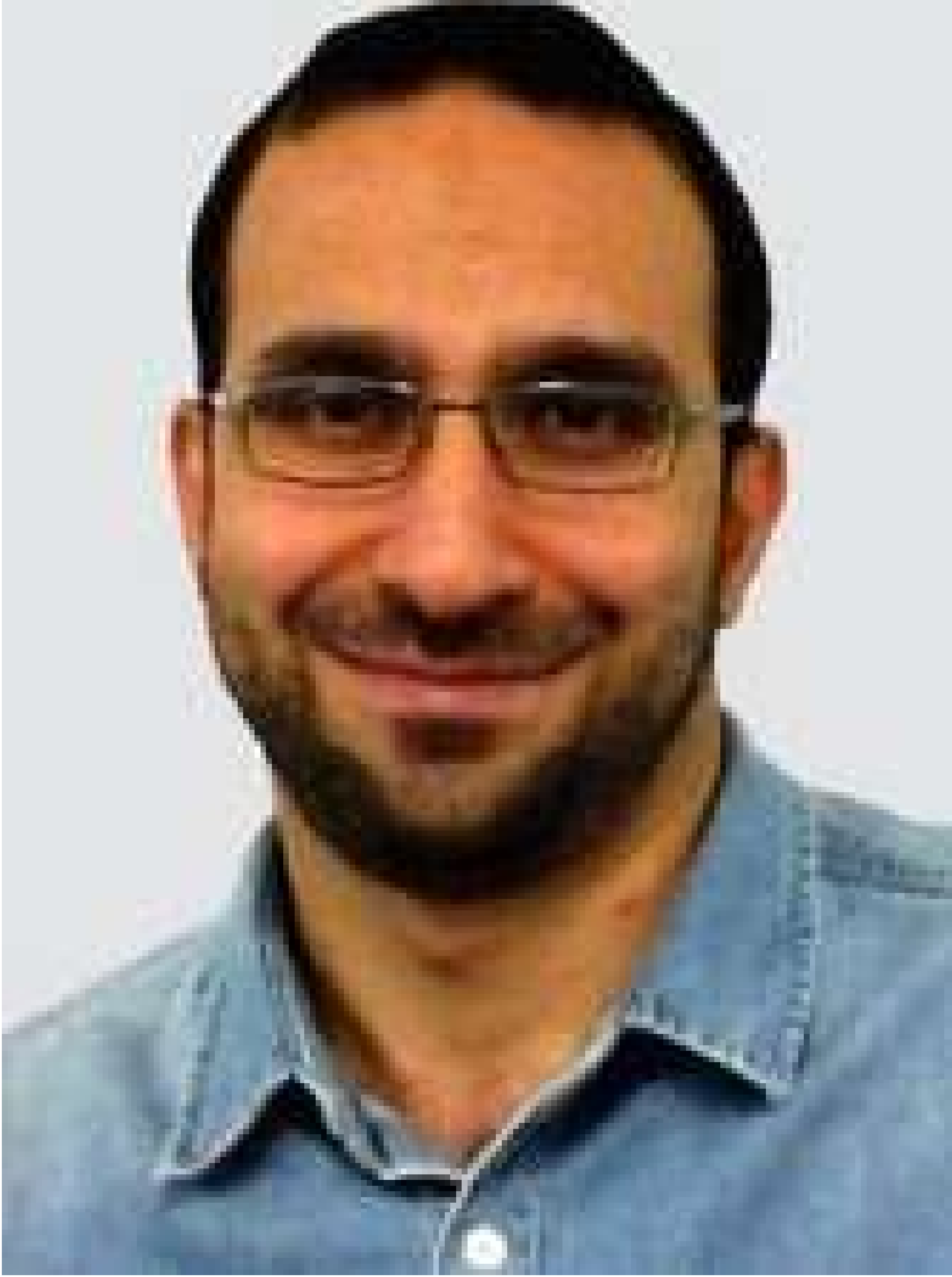}}]{Saeid Hosseini}
	won the Australian Postgraduate Award and received Ph.D. degree in Computer Science from the University of Queensland, Australia, in 2017. As a post-doc research scientist, his research interests include spatiotemporal database, dynamical processes, data and graph mining, big data analytics, recommendation systems, and machine learning. He has been a PC member in CSS and DASFAA and a reviewer in ICDM and TKDE.
\end{IEEEbiography}
\vspace{-14mm}
\begin{IEEEbiography}[{\includegraphics[width=1.0in,height=1.25in,clip,keepaspectratio]{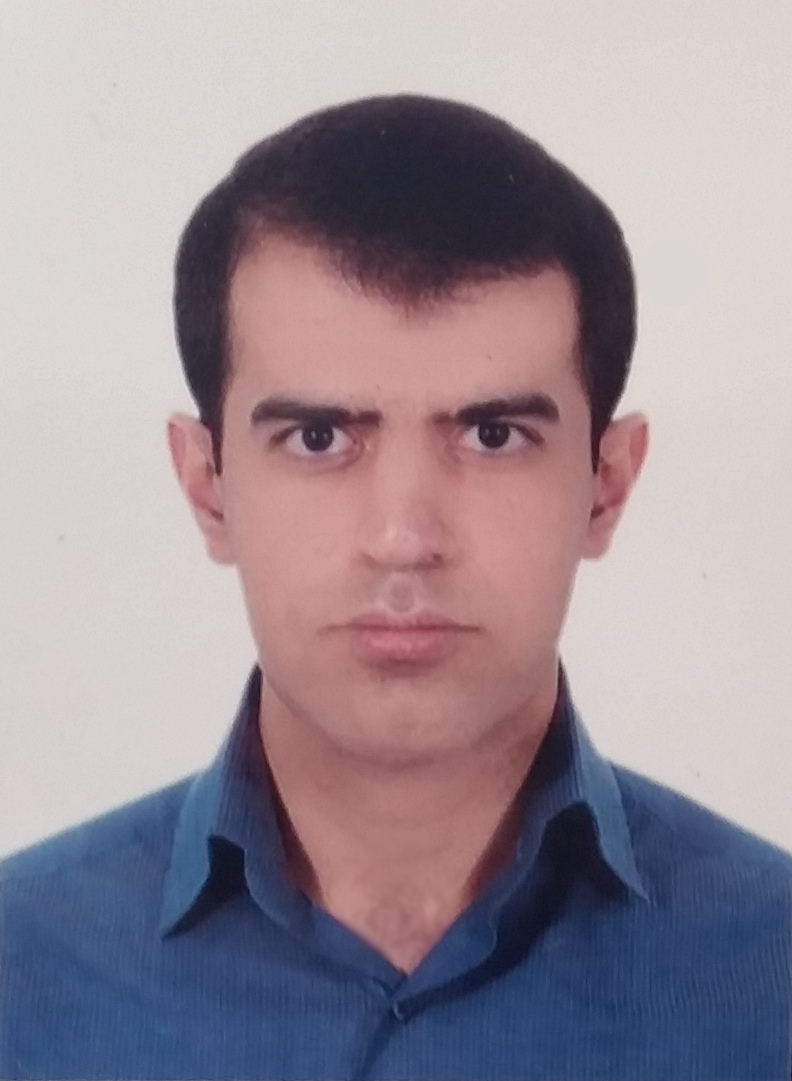}}]{Saeed NajafiPour}
	is a researcher in the computational cognitive models laboratory at Iran University of Science and Technology (IUST). He is to receive his M.Sc. in Software Engineering from IUST and completed his B.S. degree in computer software engineering from the Bahonar University, Iran. His research interests include Natural Language Processing, data mining, deep learning, and trajectory analytics.
\end{IEEEbiography}
\vspace{-14mm}
\begin{IEEEbiography}[{\includegraphics[width=1.0in,height=1.25in,clip,keepaspectratio]{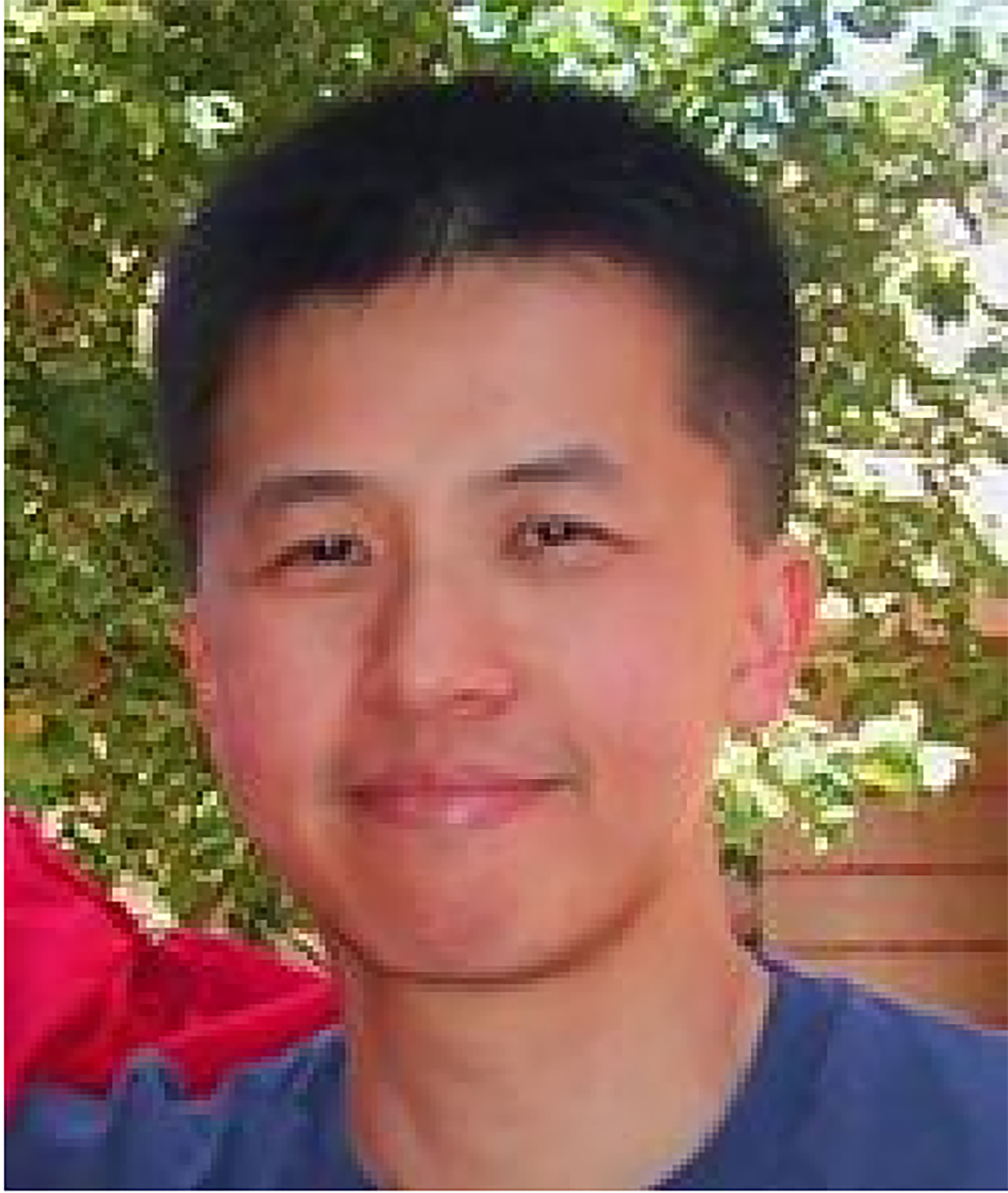}}]{Ngai-Man Cheung}
	received the Ph.D. degree in electrical engineering from
the University of Southern California, in 2008. He is currently an Assistant Professor with the Singapore University of Technology and Design (SUTD). His research interests include signal, image, and video processing, computer vision and machine learning.
\end{IEEEbiography}
\vspace{-14mm}
\begin{IEEEbiography}[{\includegraphics[width=1.0in,height=1.25in,clip,keepaspectratio]{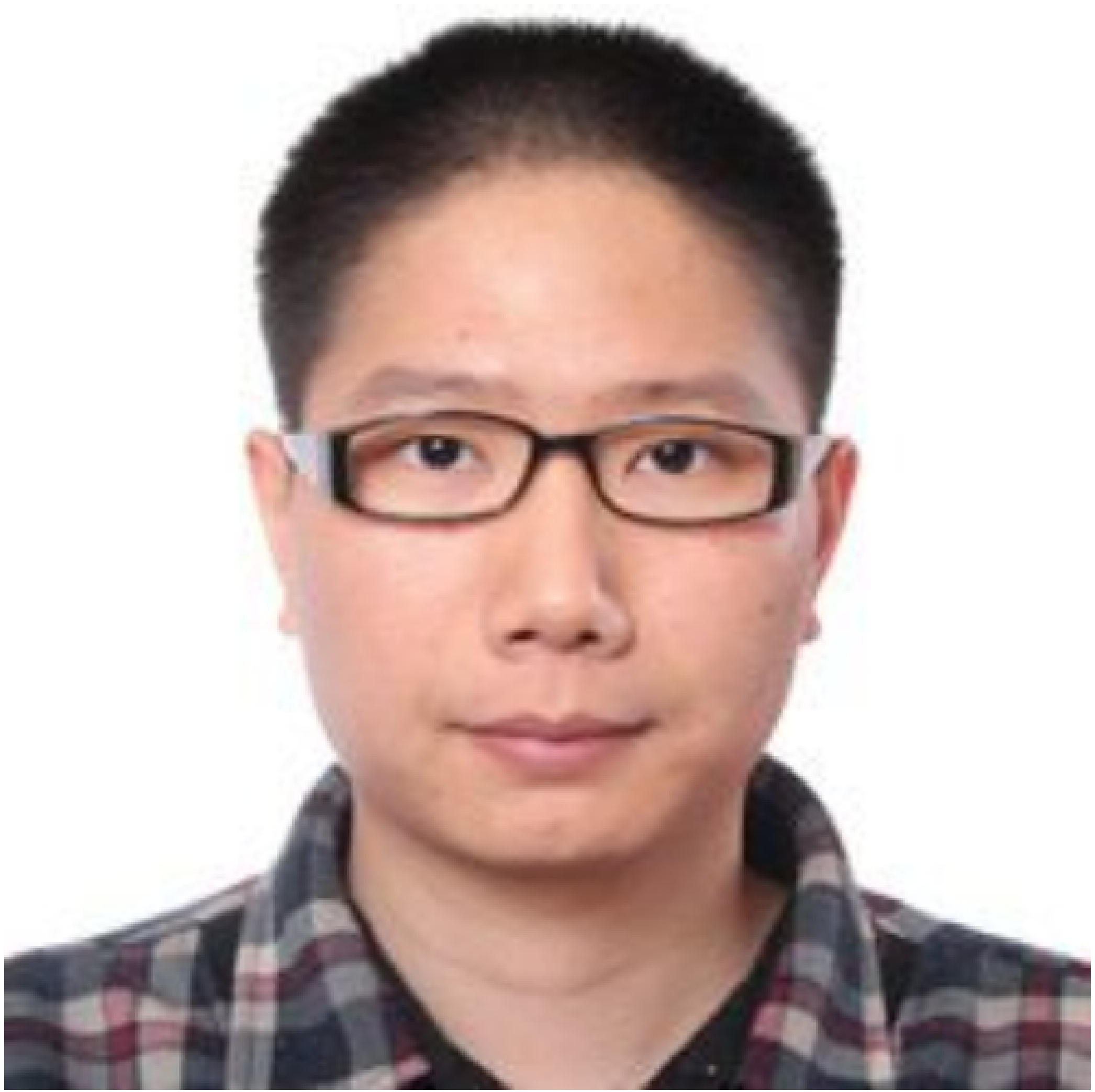}}]{Hongzhi Yin}
	works as a senior lecturer and an ARC DECRA Fellow with The University of Queensland, Australia. He has published 110+ papers and won 5 Best Paper Awards such as ICDE'19 Best Paper Award. He has published in reputed journals and top international conferences including VLDB Journal, ACM TOIS, IEEE TKDE, ACM TKDD.
\end{IEEEbiography}
\vspace{-14mm}
\begin{IEEEbiography}[{\includegraphics[width=1.0in,height=1.25in,clip,keepaspectratio]{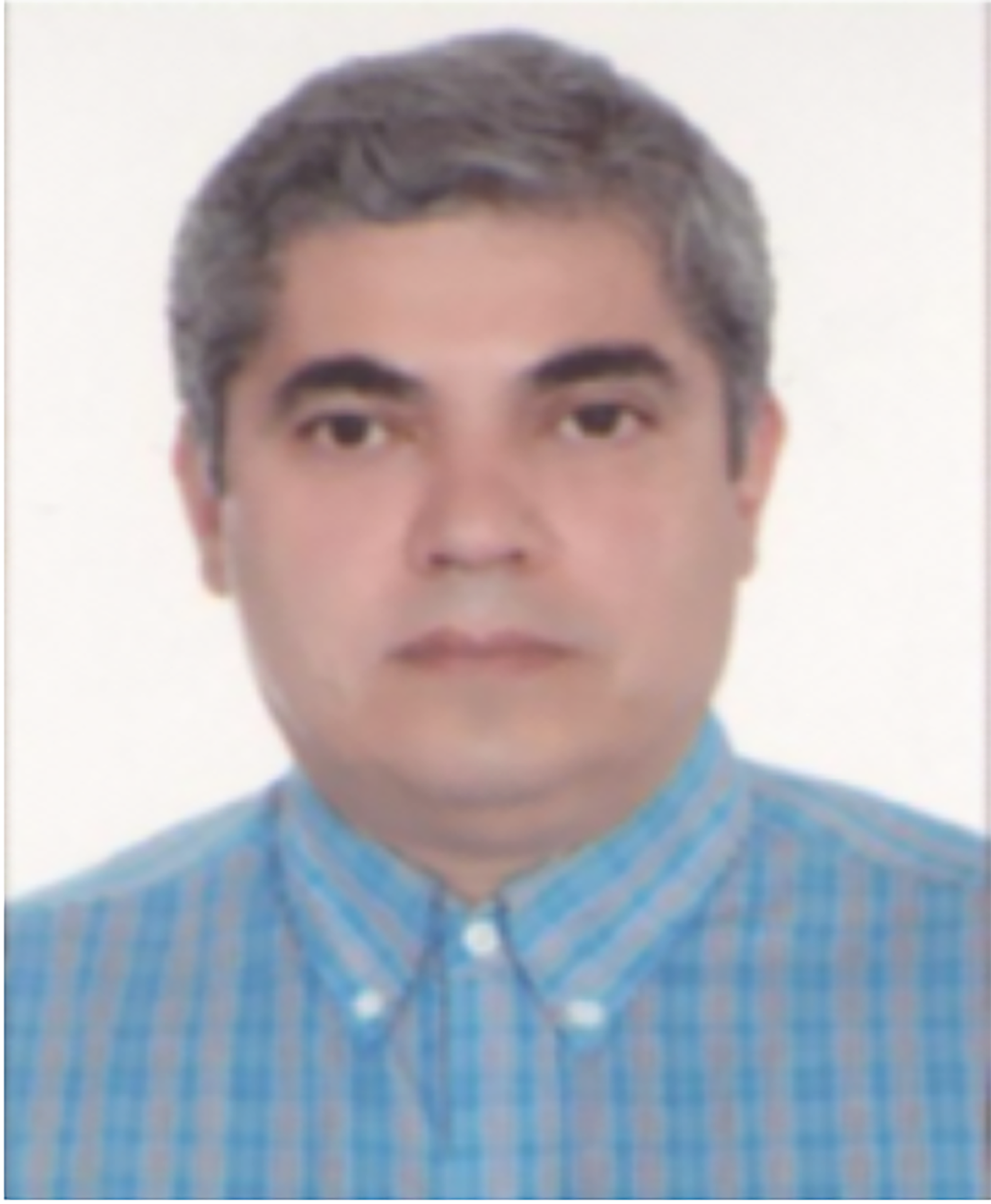}}]{Mohammad Reza Kangavari}
	received Ph.D. in computer science from the University of Manchester in 1994. He is a lecturer in the Computer Engineering Department, Iran University of Science and Technology. His research interests include Intelligent Systems, Human-Computer-Interaction, Cognitive Computing, Data Science, Machine Learning, and Sensor Networks.
\end{IEEEbiography}
\vspace{-14mm}
\begin{IEEEbiography}[{\includegraphics[width=1.0in,height=1.25in,clip,keepaspectratio]{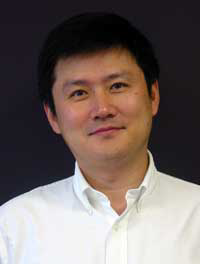}}]{Xiaofang Zhou}
	is a Professor of Computer Science at The University of Queensland, Leader of UQ Data Science Research Group and an IEEE Fellow. He received B.Sc. and M.Sc. in Computer Science from Nanjing University, China, and Ph.D. from UQ. His research interests include spatiotemporal and multimedia databases, data mining, query processing, big data analytics and machine learning, with over 300 publications in SIGMOD, VLDB, and ACM Transactions.
\end{IEEEbiography}






\end{document}